\documentclass[11pt,a4paper]{article}
\usepackage{jcappub}
\usepackage{epsfig}
\usepackage{natbib}
\usepackage{amsmath}
\usepackage{amsfonts}
\usepackage{amssymb}
\usepackage{graphicx}
\usepackage{hyperref}
\usepackage{caption}
\usepackage{subcaption}
\DeclareMathOperator{\sinc}{sinc}

\title{Self-interactions of ULDM to the rescue?}
\author[a]{Bihag Dave,}
\author[~b, c]{and Gaurav Goswami}

\affiliation[a]{School of Engineering and Applied Science, Ahmedabad University, Commerce Six Roads, Navrangpura,
Ahmedabad - 380009, Gujarat, India}
\affiliation[b]{Division of Mathematical and Physical Sciences, School of Arts and Sciences, Ahmedabad University, Commerce Six Roads, Navrangpura, Ahmedabad - 380009, Gujarat, India}
\affiliation[c]{International Centre for Space and Cosmology, Ahmedabad University,
Commerce Six Roads, Navrangpura, Ahmedabad - 380009, Gujarat, India}

\emailAdd{bihag.d@ahduni.edu.in}
\emailAdd{gaurav.goswami@ahduni.edu.in}

\abstract{One of the most important unanswered questions in cosmology is concerning the fundamental nature of dark matter (DM). DM could consist of spinless particles of very small mass i.e. $m \sim 10^{-22}$ eV. This kind of ultralight dark matter (ULDM) would form cored density profiles (called “solitons”) at the centres of galaxies. In this context, recently it has been argued that (a) there exists a power law relation between the mass of the soliton and mass of the surrounding halo called the Soliton-Halo (SH) relation, and, (b) the requirement of satisfying observed galactic rotation curves as well as SH relations is so stringent that ULDM is disfavoured from comprising $100 \%$ of the total cosmological dark matter. In this work, we revisit these constraints for ULDM particles with non-negligible quartic self-interactions. Using a recently obtained soliton-halo relation which takes into account the effect of self-interactions, we present evidence which suggests that, for $m \sim 10^{-22}\ \text{eV}$, the requirement of satisfying both galactic rotation curves as well as SH relations can be fulfilled with repulsive self-coupling $\lambda \sim \mathcal{O}(10^{-90})$.}


\arxivnumber{2304.04463}

\begin{document}
\maketitle
\flushbottom

\section{Introduction}\label{sec:Introduction}

The physical nature of Dark Matter (DM) has eluded physicists for nearly a century since its initial proposition \cite{Workman:2022ynf,Profumo:2019ujg,Safdi:2022xkm}. 
One expects DM to consist of some new kind of elementary particles, as opposed to those in the standard model of elementary particle physics. However, the basic properties of these newer elementary particles making up DM, such as spin, mass, interaction strengths etc, are completely unknown. 

If the elementary particles forming DM are fermions, Pauli's exclusion principle implies that the corresponding particle mass must be larger than ${\cal O}(100)~ {\rm eV}$ --- the well known Tremaine-Gunn bound \cite{Tremaine:1979we} --- in this picture, DM could be thought of as a collection of non-relativistic particles. On the other hand, if DM consists of Bosons, the particle masses are allowed to be smaller than this bound. In particular, for DM Bosons with masses much lighter, the occupation numbers can be so large that it can be conveniently described as a classical field \cite{Hui:2021tkt}. While this behaviour is expected for Bosons in general, to begin with, one could restrict one's attention to considering spin zero i.e. scalar particles. Most theories of new Physics do indeed predict new scalar fields which play important roles in cosmology.

Existence of DM dominated dwarf galaxies suggests that, if all particles making up DM have the same mass, the particle mass can't be too small i.e. $m \ge 10^{-22} ~ {\rm eV}$ \cite{Hu:2000ke, Matos:1999et, Matos:2000ss}.
The term Fuzzy Dark Matter (FDM) is used to refer to
Ultra-Light Dark Matter (ULDM) particles in the mass range $m \sim 10^{-22} - 10^{-20}~ {\rm eV}$ with negligible self-interactions~\cite{Urena-Lopez:2019kud, Niemeyer:2019aqm, Ferreira:2020fam}. An important feature of FDM is the core-halo structure, predicted by various independent simulations \cite{Schive:2014dra, Schive:2014hza, Mocz:2017wlg, Mina:2020eik} over the last decade. The inner regions of DM halos are described by flat density cores (or solitons) while far from the centre, the density transitions to a CDM-like profile, essentially not altering large scale cosmology. Due to this, FDM can solve the `core-cusp' problem, the missing satellites problem, and the too-big-to-fail problem \cite{Bullock:2017xww}.
The simulations which suggest that FDM solves these problems also suggest a power law relation between the mass of the soliton and mass of the surrounding halo called the Soliton-Halo (SH) relation. Note that there is some disagreement on value of the exponent in the power law \cite{Chan:2021bja}. 

Recently, Refs.~\cite{Bar:2018acw, Bar:2021kti} have argued that FDM in the mass range $m \in \left[10^{-24}\ \text{eV}, 10^{-20}\ \text{eV}\right]$ cannot adequately describe observed rotation curves from the Spitzer Photometry \& Accurate Rotation Curves (SPARC) catalogue while also satisfying the Soliton-Halo (SH) relation obtained by \cite{Schive:2014dra, Schive:2014hza}. This along with other concerns about FDM~\cite{Robles:2018fur, Hlozek:2017zzf, Lague:2021frh, Irsic:2017yje, Kobayashi:2017jcf, Rogers:2020ltq, Marsh:2018zyw, Safarzadeh:2019sre, Banares-Hernandez:2023axy} suggest that the assumptions involved in the FDM paradigm need to be carefully examined.

Fundamental physics suggests that scalars should have self coupling, in particular, the quartic self coupling. While the self-coupling of the scalar particles forming DM could indeed be negligible, whether this is the case needs to be established by observations \cite{Li:2013nal, Suarez:2016eez, Desjacques:2017fmf, Cembranos:2018ulm}. It is thus natural to consider DM consisting of ultra-light classical scalar fields with non-negligible self interactions. To aid the discussion, let us refer to ULDM consisting of a classical scalar field with attractive or repulsive self-interactions by the name scalar field dark matter (SFDM).

To model cores of DM halos, self gravitating configurations of such self interacting scalar fields have to be considered in great detail. One can form stable configurations of different masses and sizes (depending on the sign and strength of self-interactions), such as Boson stars, Q-balls, Oscillatons, etc. \cite{Visinelli:2021uve}. It is well known that even a very small self-coupling can dramatically affect the structure and stability of the resulting pseudo-solitonic solutions \cite{Colpi:1986psi, Chavanis:2011mrr}. Given this, it is important to find out if SFDM with attractive or repulsive self coupling can help evade the constraints obtained by Refs.~\cite{Bar:2018acw, Bar:2021kti} based on rotation curves and soliton-halo relations. In order to answer this question, it is also important to develop a clear understanding of the behaviour of solitonic cores in the presence of arbitrary self interactions of the scalar field. We explore these issues in the present paper.

This paper is organised as follows: in section~\ref{sec:observations}, we briefly discuss rotation curves, SH relations and the results of \cite{Bar:2021kti}. We also summarize the motivation for including self-interactions, SH relations in the presence of self-interactions, and introduce the Gross-Pitaevskii-Poisson system along with its time-independent solutions. In section~\ref{sec:mass_radius_curves}, we revisit the so-called mass-radius relations for attractive and repulsive self-interactions using numerical solutions and scaling symmetry, while in section~\ref{sec:impact_of_parameters} we examine the impact of self-interactions on rotation curves of solitons. In section~\ref{sec:analysis_SPARC}, using a modified SH relation, we show that SFDM with $\lambda > 0$ can simultaneously satisfy the said relation and is allowed by the data. In section~\ref{sec:summary} we summarize our results and motivate future work from our initial analysis. In appendix~\ref{app:extension_m_r_curves}, we describe the various regimes in the mass-radius plane for a fixed $\lambda$, address a constraint from our previous work \cite{Chakrabarti:2022owq} for $\lambda < 0$ and show how the mass-radius curve is altered in the presence of a black hole at the centre. Recently it was shown that the SH relation for FDM can be written as the condition that peak velocities in the soliton and halo be approximately equal \cite{Bar:2018acw}. In appendix~\ref{app:slope_peak_condition} we impose this condition while also allowing for attractive and repulsive self-interactions for a fixed $m = 10^{-22}\ \text{eV}$ and confront DM-only velocity curves of 17 LSB galaxies.

We work in $\hbar = c = 1$ units unless mentioned otherwise, and denote Planck mass and reduced Planck mass by  $m_{pl}$ and $M_{pl}$ respectively.

\section{Observations and Motivation}\label{sec:observations}

Galactic rotation curves i.e. orbital velocity of stars and gas as a function of distance from the centres of galaxies are an important probe of the matter (visible and dark) distribution in said galaxies \cite{Carroll:2017ima, Sofue:2000jx}. In general one can obtain the circular velocity of a test particle in an orbit of radius $r$ in the gravitational potential of a spherically symmetric distribution of matter using

\begin{equation}\label{eq:circ_vel}
v(r) = \sqrt{\frac{GM(r)}{r}} = \sqrt{\frac{4\pi G\int_0^r\rho(r')r'^2dr'}{r}}\ .
\end{equation}
The total observed velocity can be split into various components corresponding to different distributions of matter in the galaxy: (a) stellar disk ($V_d$), (b) stellar bulge ($V_b$), (c) gas ($V_g$) and (d) dark matter ($V_{DM}$). Hence, a typical observed rotation curve for a galaxy from, for instance the Spitzer Photometry \& Accurate Rotation Curves (SPARC) catalogue \cite{Lelli:2016zqa}, can be written as

\begin{equation}\label{eq:observed_vel}
    V_{obs} = \sqrt{V_{DM}^2 + V_{g}|V_{g}| + \Upsilon_d V_{d}|V_{d}| + \Upsilon_b V_{b}|V_{b}|}\ .
\end{equation}
Note that here contributions from the disk and bulge can be tuned using the stellar mass-to-light ratios $\Upsilon_d$ and $\Upsilon_b$ respectively. Usually, observed rotation curves exhibit a velocity that increases in the inner region, and then flattens as one goes further away from the centre. The inner regions of large galaxies are well-explained by the sizeable amount of baryonic matter contained near the centre. On the other hand, dark matter is required to explain the flat rotation curves at large $r$ where baryonic contribution is very little. DM-only simulations \cite{Navarro:1995iw} suggest a density profile for DM that goes like $r^{-1}$ in the inner region and $r^{-3}$ at large $r$. This is the well-known Navarro-Frenk-White (NFW) profile given by

\begin{equation}
    \rho_{NFW}(r) = \frac{\rho_s}{\frac{r}{r_s}\left(1 + \frac{r}{r_s}\right)^2}\ ,
\end{equation}
where $\rho_s$ and $r_s$ are parameters of the profile. These parameters can be chosen such that the corresponding velocity curve exhibits a flat portion at a desired scale. The NFW velocity curve also attains a maximum at $\sim 2.16r_s$. For small $r$, using eq.~(\ref{eq:circ_vel}) one can see that NFW profile implies an increasing velocity where $v\propto \sqrt{r}$. However, for many low mass and low surface brightness (LSB) galaxies where the baryonic contribution is thought to be small even at small radius, observed velocities in the inner regions ($\sim \mathcal{O}(1)\ \text{kpc}$) point to a more slowly increasing velocity curve, $v \propto r$. This is the manifestation of the well-known core-cusp problem \cite{Bullock:2017xww}. 

FDM resolves this issue by considering the inner regions to be described by stable solutions of the Schrodinger-Poisson equations. These core-like structures have flat density profiles and are called solitons. Independent numerical simulations \cite{Schive:2014dra, Schive:2014hza, Mocz:2017wlg, Mina:2020eik} have confirmed a core-halo structure for FDM, where the inner region is described by the FDM core (also called a soliton) while further from the centre, it behaves like CDM. Here, the total density profile can be written as, 

\begin{equation}\label{eq:tot_dens}
    \rho(r) = \Theta{(r_t - r)}\rho_{SFDM}(r) + \Theta{(r - r_t)}\rho_{NFW}(r)\ ,
\end{equation}
where imposing the continuity of density at $r_t$ implies $\rho_{SFDM}(r_t) = \rho_{NFW}(r_t)$. Note that this fixes one of the parameters of the NFW profile ($\rho_s$) leaving two free parameters for the outer envelope: $\{r_t, r_s\}$.

\subsection{Soliton-Halo (SH) relations}\label{sec:SH_relations_obs_curves}

Simulations in \cite{Schive:2014dra, Schive:2014hza} also obtained a power-law relationship between mass of the soliton $M_{SH}$ and mass of the halo $M_h$\footnote{We remind the reader that halo mass is often defined as $M_{h} = \frac{4\pi}{3}(200\rho_{c})R_{200}^3$, where $R_{200}$ is the radius at which the average density of the mass contained is $200$ times the critical density ($\rho_c$) of the universe.} of the form $M_{SH} \propto M_h^{1/3}$, or, more precisely \cite{Schive:2014hza}

\begin{equation}\label{eq:Schive_SH}
\left(\frac{M_{SH} }{10^9\ M_\odot}\right) = 1.4\left(\frac{M_h}{10^{12}\ M_\odot}\right)^{1/3}\left(\frac{m}{10^{-22}\ \text{eV}}\right)^{-1}\ .
\end{equation}
The simulations report a scatter of roughly a factor of $2$ in this soliton-halo (SH) relation. It is important to note that other simulations \cite{Mocz:2017wlg, Mina:2020eik} have reported a different power law between soliton mass and halo mass: $M_{SH} \propto M_h^{0.556}$: this disagreement could partly be the result of different merger histories and tidal stripping \cite{Chan:2021bja}. 

Therefore, if simulations in \cite{Schive:2014dra, Schive:2014hza} are correct, then along with describing observed rotation curves, FDM is expected to satisfy the SH relation in eq.~(\ref{eq:Schive_SH}). In other words, such relations can be considered to be a sharp prediction of the FDM paradigm.

\subsection{Rotation curves and SH relations for FDM}\label{sec:FDM_constraints}

Recently, using all 175 galaxies in the SPARC catalogue, Ref.~\cite{Bar:2021kti} reported that, for FDM, the SH relation in eq.~(\ref{eq:Schive_SH}) is not consistent with the observed rotation curves. 
In this context, we direct the reader to figure~1 of Ref.~\cite{Bar:2021kti} as well as figure~\ref{fig:LSB_no_inter} of this work. 
The details of the procedure followed in Ref.~\cite{Bar:2021kti} which are relevant for our purpose are discussed in section~\ref{sec:imposing_mod_SH_relations}. For now, we just highlight the following: the soliton masses allowed by the rotation curves data were much smaller than the corresponding soliton masses expected from the SH relation obtained from eq.~(\ref{eq:Schive_SH}) for $m\in \left[10^{-24}\ \text{eV}, 10^{-20}\ \text{eV}\right]$.\footnote{Authors in \cite{Bar:2019bqz} found that the SH relation in eq.~(\ref{eq:Schive_SH}) was equivalent to the ratio of kinetic energy and total mass being roughly the same for the soliton and halo: $(K/M)_{sol} \approx (K/M)_{halo}$. In particular, Ref.~\cite{Bar:2018acw} showed that for $m\in\left[10^{-22}\ \text{eV}, 10^{-21}\ \text{eV}\right]$, if a soliton with mass $M_s$ is expected to satisfy the SH relation, then the corresponding velocity curve significantly overshoots observed velocity in the inner regions of dark matter dominated galaxies.}

Hence, the SH relation for an ultra-light scalar field with no self-interactions seems to be incompatible with observed rotation curves. This along with other constraints mentioned in section~\ref{sec:Introduction} potentially rules out FDM being a significant fraction of all dark matter.

\subsection{SFDM self-interactions}
\label{sec:selfinter}

Let us begin by noting that the existence of quartic (i.e. $\lambda\varphi^4$-type) self-interaction term in the Lagrangian of a scalar field is inevitable. In the non-relativistic limit (relevant to cold DM), this self-interaction leads to an inter-particle interaction potential energy function of the form $U = U_0 ~ \delta^3 ({\bf x}_i - {\bf x}_j)$ i.e. it is a contact interaction. Depending on the sign of self coupling, this interaction could be attractive ($\lambda < 0$) or repulsive ($\lambda > 0$). 

Scalar field dark matter (SFDM) with attractive self-interactions is well motivated if one considers axions, where Taylor expansion of the cosine potential will lead to a quartic self-interaction term with $\lambda < 0$ \cite{Marsh:2015xka}. On the other hand, repulsive ($\lambda > 0$) self-interactions are expected from e.g. moduli fields ubiquitous in theories of high energy physics. 

If ultra-light axions (ULAs) are to comprise all of DM, the self-interaction strength must be of the order $\sim 10^{-96}$ for mass $\sim 10^{-22}\ \text{eV}$ \cite{Hui:2016ltb}. Recent simulations in \cite{Mocz:2023adf} focus on the impact of such attractive self-interactions on cosmological structure formation. In fact, it is well known that even very small self-interactions can dramatically change the resultant stable configuration \cite{Colpi:1986psi}. The effect of self-interactions on the mass and radius of solitons is apparent even in the Newtonian-limit, as \cite{Chavanis:2011mrr} demonstrated for both attractive and repulsive self-interactions. 

Thus, there are very good reasons to consider SFDM with small but non-negligible self-interactions. Before proceeding, we note that SFDM with self interactions, in particular in the Thomas-Fermi (TF) regime \cite{Boehmer:2007um}, has been constrained in the past in various ways e.g. by looking at cosmological evolution and structure formation \cite{Li:2013nal, Foidl:2022bpn}, and even using rotation curves \cite{Bernal:2017oih, Craciun:2020twu, Harko:2022itw, Dawoodbhoy:2021beb, Delgado:2022vnt}. See also \cite{Khlopov:1985jw, Dev:2016hxv, Banerjee:2022zii, Davoudiasl:2023uiq, Fox:2023aat, Feng:2021qkj} for some selected references that consider scalar fields with self-interactions in this context.

\subsection{SH relations for SFDM with self-coupling}\label{sec:SFDM_SH_relations}

Other parameters being fixed, the mass of a soliton in the presence of self-interactions gets changed. 
This suggests that, in the presence of self-interactions, the corresponding soliton-halo relation will take up a form different from eq.~(\ref{eq:Schive_SH}) which is not expected to be valid when $\lambda \neq 0$. Recently, \cite{Chavanis:2019faf, Padilla:2021chr} arrived at the corresponding SH relation, which takes up the following form (see also section~V of Ref.~\cite{Chavanis:2020rdo})

\begin{equation}\label{eq:modified_SH}
\left(\frac{M_{SH} }{10^9\ M_\odot}\right) = 1.4\left(\frac{M_h}{10^{12}\ M_\odot}\right)^{1/3}\left(\frac{m}{10^{-22}\ \text{eV}}\right)^{-1}\sqrt{1 + (1.16\times 10^{-7})\hat{\Lambda}\left(\frac{M_h}{10^{12}\ M_\odot}\right)^{2/3}}\ ,
\end{equation}
where the $\hat{\Lambda}$ in the above equation is proportional to the self coupling $\lambda$ of the scalar.\footnote{
More precisely, $\hat{\Lambda}$ is the same as $\frac{\lambda}{4} \left( \frac{M_{pl}}{m} \right)^2$ which is the same as $2 (s^2\hat{\lambda}_{ini})$ in our notation introduced in section~\ref{sec:GPP_system}.} The origin of the numerical factor in front of $\hat{\Lambda}$ can be understood from the discussion above eq.~(80) in \cite{Padilla:2021chr}. Note that in the absence of self-interactions the SH relation reduces to eq.~(\ref{eq:Schive_SH}).
While eq.~(\ref{eq:modified_SH}) is valid for both attractive and repulsive self-interactions, for $\lambda$ that is too negative, this SH relation will no longer be applicable (see also section~\ref{sec:why_no_att_inter}).

\subsection{Gross-Pitaevskii-Poisson  Equations}\label{sec:GPP_system}

Consider a classical real scalar field with the potential $U(\varphi) = \frac{m^2\varphi^2}{2} + \frac{\lambda\varphi^4}{4!}$, where, $\lambda$ dictates the strength of self-interactions. In the non-relativistic limit, the real $\varphi$ can be written in terms of a complex field, $\varphi = \frac{1}{\sqrt{2}m}\left(e^{-imt}\Psi + c.c.\right)$. By averaging out the rapidly oscillating modes and taking the weak-gravity limit, the scalar field can then be described using the Gross-Pitaevskii-Poisson (GPP) equations \cite{Pitaevskii:2016book}
(see \cite{Chakrabarti:2022owq} for a detailed derivation and other notations and conventions used in this paper):

\begin{eqnarray} 
     i \frac{\partial \Psi}{\partial t} &=&  -\frac{\nabla^2}{2m} \Psi + m \Phi  \Psi +  \frac{\lambda}{8 m^3} |\Psi|^2 \Psi \ , \label{eq:GrossPitaevskii} \\
    \nabla^2 \Phi &=& \frac{|\Psi|^2}{ 2 M_{\rm pl}^2 } \ . \label{eq:Poisson}
\end{eqnarray}
Here the mass density of the scalar field is $\rho = |\Psi|^2$. For modeling cores of DM halos, we are interested in stationary solutions of the GPP system such that $\Phi$ is time-independent and one can separate time dependence for $\Psi$ using $\Psi(\vec{r}, t) = \phi(\vec{r})e^{-i\gamma t}$. We also want solutions to be spherically symmetric, node-less (ground state), spatially localised and regular everywhere: in the present context, such solutions are referred to as solitons. 

\begin{figure}[ht]
    \centering
    \includegraphics[width = 0.6\textwidth]{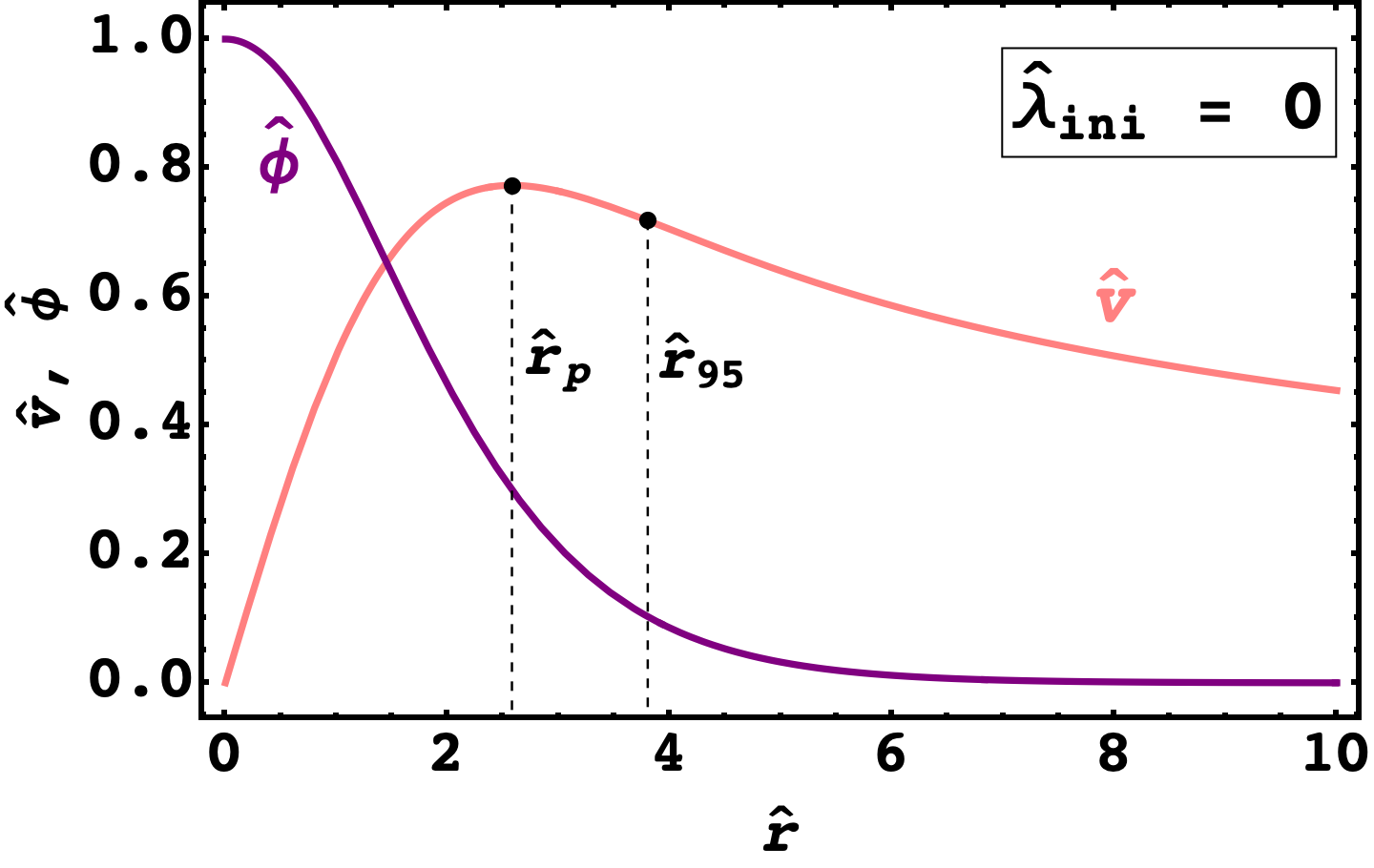}
    \caption{Dimensionless solution satisfying the boundary condition $\hat{\phi}(\infty) = 0$ for $\hat{\lambda}_{ini} = 0$ is shown by the purple curve, while the corresponding velocity curve is shown by the pink curve. Note that $\hat{r}_{p}$ is the radius at which the velocity peaks.}
    \label{fig:density_velocity}
\end{figure}

Even if the microscopic parameters of DM species e.g. the particle mass $m$ and self coupling $\lambda$ are fixed, one can form many possible solitons (which are macroscopic objects) depending on how many particles form the soliton. 
In other words, even for fixed $m$ and $\lambda$, there is a family of solutions parameterised by the total number of particles used to form them. It is convenient to work with dimensionless variables which motivates the following re-scaling of relevant quantities: $\hat{\phi} = \frac{\hbar\sqrt{4\pi G}}{mc^2}\phi$, $\hat{\Phi} = \frac{\Phi}{c^2}$, $\hat{\gamma}  = \frac{\gamma}{mc^2}$, $\hat{\lambda} = \frac{\lambda}{8}\left(\frac{M_{pl}}{m}\right)^2$, $\hat{r} = \frac{m c r} {\hbar}$. Note that $m$ will be absent in the dimensionless GPP equations. For the rest of the paper, any `hatted' variable will denote a dimensionless quantity. One can then use the shooting method to solve the system with the following initial conditions: $\hat{\phi}(0) = 1$, $\hat{\phi}'(0) = 0$, $\hat{\Phi}(0) = 0$, $\hat{\Phi}'(0) = 0$. One can obtain the value of $\hat{\gamma}$ which satisfies the boundary condition $\hat{\phi}(\hat{r}\rightarrow\infty) = 0$ (solution for $\hat{\lambda} = 0$ is shown by the purple curve in figure~\ref{fig:density_velocity}). 

Note that $\hat{\lambda}$ is the sole free parameter of the dimensionless GPP system. To obtain the family of solutions mentioned above, one can get to a physical solution with the desired value of central density $\hat{\phi}(0)$ using the scaling symmetry that the GPP system enjoys \cite{Davies:2019wgi, Guzman:2006yc, Li:2020ryg}:

\begin{equation}\label{eq:scaling}
    \left\{r, \lambda, \phi, \Phi, \gamma\right\} \rightarrow  \left\{sr, s^2\lambda, s^{-2}\phi, s^{-2}\Phi,  s^{-2}\gamma\right\}\ ,
\end{equation}
where $s$ is the scaling parameter (see \cite{Chakrabarti:2022owq} for details).
The value of the scaling parameter $s$ is the numerical factor by which the soliton becomes larger or smaller due to scaling transformation. Taking into account this way of parameterizing the solutions, free parameters of the dimensionful GPP system then are $\{m, \hat{\lambda}_{ini}, s\}$, where ${\hat \lambda}_{ini}$ is the value of ${\hat \lambda}$ before scaling.

Once a solution and the corresponding density profile is obtained, one can also calculate the total soliton mass by solving the integral,

\begin{equation}\label{eq:Mhat}
    \hat{M} = \int_0^\infty \hat{r}^2\hat{\phi}^2d\hat{r}\ .
\end{equation}
Dimensions can be restored to obtain the physical soliton mass using $M_s = \hat{M}\frac{m_{pl}^2}{m}$. At this stage, we need to distinguish between the soliton mass $M_s$ obtained here and the soliton mass $M_{SH}$ obtained from soliton halo relation such as eq. (\ref{eq:modified_SH}). If soliton halo relations are satisfied, these two must be equal (to within the scatter of the relation). On the other hand, if the soliton halo relations do not get satisfied, $M_s$ and $M_{SH}$ could be completely different.

One can also define a characteristic length scale $\hat{R} \equiv \hat{R}_{95}$, which is the radius within which $95\%$ of the total mass is contained (similarly one can also use $\hat{R}_{99}$). Scaling for derived quantities like $\hat{M}$ can be obtained from eq.~(\ref{eq:Mhat}) and (\ref{eq:scaling}) to be $\hat{M} \rightarrow \hat{M}/s$. For the rest of the paper, we denote unscaled quantities with a subscript `ini', e.g., $\hat{\lambda}_{ini}$, $\hat{M}_{ini}$, $\hat{R}_{ini}$, etc.  

Once density profile $\hat{\rho}(\hat{r}) = |\hat{\phi}(\hat{r})|^2$ is obtained from the solution, one can also find the corresponding velocity curve from eq~(\ref{eq:circ_vel}). It is easy to see that $\hat{v} = v/c$ (velocity curve for $\hat{\lambda}_{ini} = 0$ is shown by the pink curve in figure~\ref{fig:density_velocity}). Note that scaling symmetry implies that velocity scales as $\hat{v}\rightarrow \hat{v}/s$. 

\section{Self-interacting ULDM, rotation curves and soliton-halo relations}\label{sec:soliton_properties}
In this section we present our main results. 

Before proceeding, note that we fix DM particle mass to the fiducial value of $m = 10^{-22}\ \text{eV}$ unless mentioned otherwise and hence focus our attention on the effect of (a) varying DM self-coupling $\lambda$ (parameterised by the dimensionless quantity ${\hat \lambda}_{ini}$), and, (b) varying the total number of DM particles forming the soliton acting as the core of the DM halo of a given galaxy (this is parameterised by soliton mass $M_s$ or scale $s$). 

For a particular DM species with a fixed physical $m$ and $\lambda$, as we consider various soliton solutions with different total masses, the size of the soliton is different. In section~\ref{sec:mass_radius_curves} we arrive at some important results about the connection between the mass of the soliton and its radius. 
Since we eventually need to satisfy rotation curves, in section~\ref{sec:impact_of_parameters} we shall briefly look at the impact of the free parameters $\{\hat{\lambda}_{ini}, s\}$ on the circular velocity (i.e. rotation curves). Finally, in section~\ref{sec:imposing_mod_SH_relations} we shall check the compatibility of the modified SH relation in eq.~(\ref{eq:modified_SH}) with observed rotation curves for solitons formed from SFDM with self-interactions.

\subsection{Mass-Radius relations and their implications}\label{sec:mass_radius_curves}

Solitonic solutions are the result of a delicate balance between the outward `quantum pressure' arising from the gradient term in the Gross-Pitaevskii equation, the attractive or repulsive self-interactions of the scalar field, and its self-gravity, leading to a family of allowed masses $M$ and corresponding sizes $R$. Even small values of the self-coupling strength, $\lambda \sim \mathcal{O}(10^{-98})$ can impact the allowed mass and size of solitons.\footnote{This can be easily seen from eq.~(\ref{eq:GrossPitaevskii}), where the self-interaction term is comparable to other terms only when $\hat{\lambda}\sim \mathcal{O}(1)$, which for $m\sim 10^{-22}\ \text{eV}$, gives the self-coupling strength $\lambda \sim \mathcal{O}(10^{-98})$.}

To understand the relation between mass and radius of solitonic solutions, one usually proceeds by using an ansatz \cite{Chavanis:2011mrr, Schiappacasse:2017ham} for the form of density profile --- this allows one to write an analytical expression for the energy of the system in terms the soliton mass $M$ and soliton radius $R$. The solutions then correspond to the critical points of the energy. One can then obtain an analytical mass-radius relation for various critical solutions \cite{Chavanis:2011mrr, Schiappacasse:2017ham}. 
This approach, while beautiful, relies on assuming a form of the density profile ansatz. It is thus interesting to ask how the expected relationship between the size of the soliton and its mass arises from numerically solving the dimensionless GPP equations. We show that the scaling symmetry of GPP system can be exploited to understand this. 

\begin{figure}[ht]
    \centering
    \includegraphics[width = 0.85\textwidth]{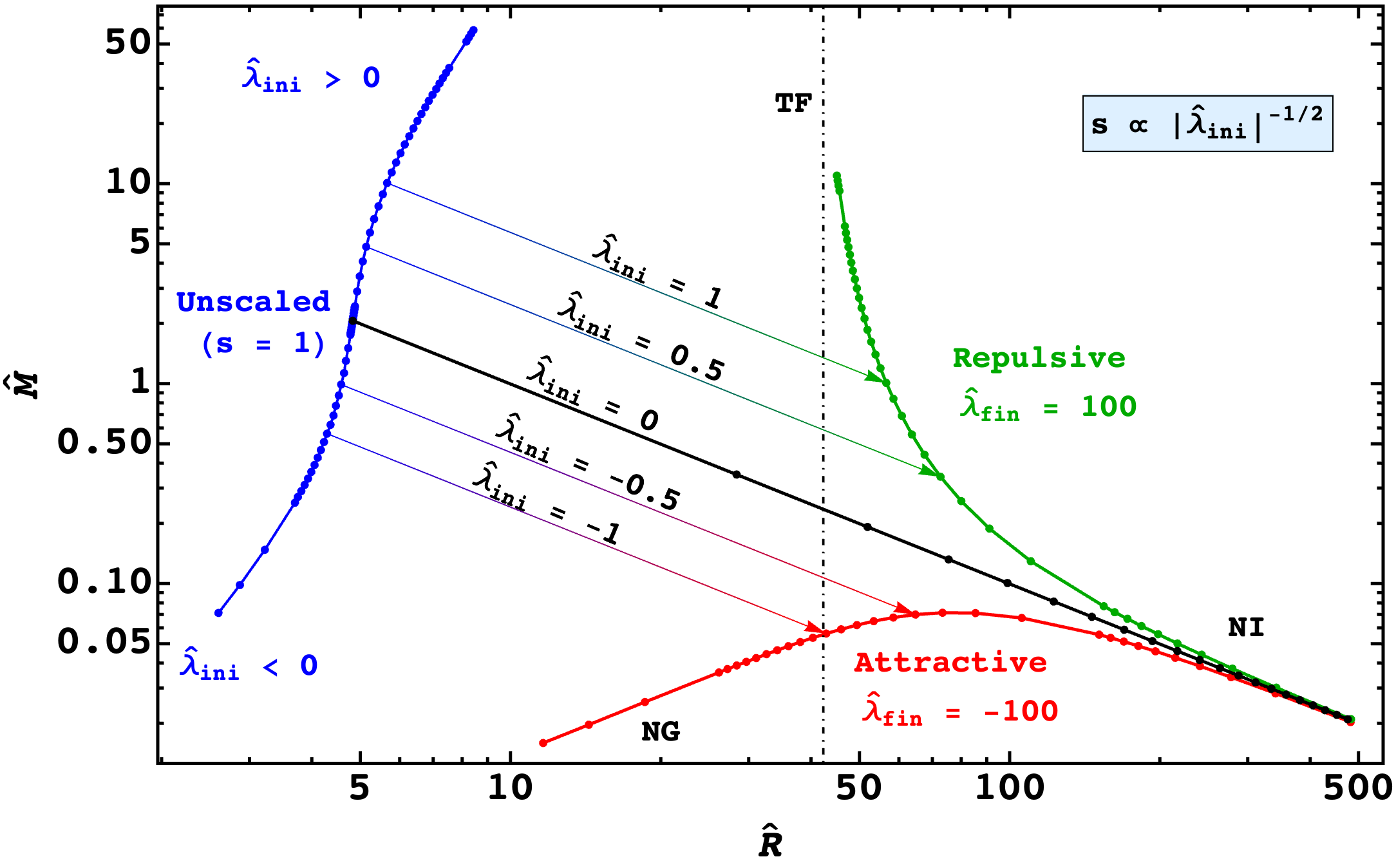}
    \caption{Blue curve denotes unscaled $\left(\hat{M}_{ini}, \hat{R}_{ini}\right)$ for various values of $\hat{\lambda}_{ini}$. Red curve shows the mass-radius curve for attractive self-interactions, while the green curve does the same for repulsive self-interactions, for a fixed $|\hat{\lambda}_{fin}| = 100$. Arrows denote transformation due to scaling from a fixed $s$ to a fixed $\hat{\lambda}_{fin}$ curve. NG corresponds to the non-gravitational regime, NI to the non-interaction regime, and TF (vertical dashed line) to the Thomas-Fermi regime. See appendix~\ref{app:extension_m_r_curves} for a discussion on different regimes.}
    \label{fig:m_vs_r_scaled}
\end{figure}

\subsubsection{Mass-radius curves without a density profile ansatz}\label{sec:m_vs_r_procedure}

We begin by solving the GPP system for various choices of $\hat{\lambda}_{ini}$, and for each such choice, we calculate the corresponding soliton mass $\hat{M}_{ini}$ and radius $\hat{R}_{ini} = \hat{R}_{99}$ (see blue curve in figure~\ref{fig:m_vs_r_scaled} marked ``Unscaled"). All the other solutions of GPP system can be obtained from this blue curve by employing scaling transformations as we now argue.

The mass-radius curves for some fixed value of scaled self-interaction strength $|\hat{\lambda}_{fin}|$ are shown in figure~\ref{fig:m_vs_r_scaled} for both attractive (red) and repulsive (green) cases. Let us now understand how one gets these curves from the unscaled blue curve. 

For some fixed arbitrary scale value $s$, using eq.~(\ref{eq:scaling}) each point can be scaled to $\left(\hat{R}_{fin}, \hat{M}_{fin}\right) = \left(s\hat{R}_{ini}, \hat{M}_{ini}/s\right)$. However, scaling symmetry also implies that each scaled point corresponds to a different value of the scaled self-interaction strength ($\hat{\lambda}_{fin} = s^2\hat{\lambda}_{ini}$). Since we are interested in the case where $\hat{\lambda}_{fin}$ (negative or positive) is fixed, we choose $s$ such that for any $\hat{\lambda}_{ini}$,

\begin{equation}\label{eq:lambda_scale}
    s = \sqrt{\frac{\hat{\lambda}_{fin}}{\hat{\lambda}_{ini}}}\ ,
\end{equation}
where $\hat{\lambda}_{fin}$ remains fixed. The corresponding scaled radius and mass of the soliton $\left(\hat{R}_{fin}, \hat{M}_{fin}\right)$ represent the mass-radius curves for a fixed $\hat{\lambda}_{fin}$ as shown in figure~\ref{fig:m_vs_r_scaled} for $\hat{\lambda}_{fin} = +100$ (green curve) and $\hat{\lambda}_{fin} = -100$ (red curve). Note from eqs.~(\ref{eq:scaling}) and~(\ref{eq:lambda_scale}) that the scaled mass $\hat{M}_{fin}$ and radius $\hat{R}_{fin}$ change only when the products $\hat{M}_{ini}|\hat{\lambda}_{ini}|^{1/2}$ and $\hat{R}_{ini}|\hat{\lambda}_{ini}|^{-1/2}$ vary respectively. This enables one to go from solutions with different $\hat{\lambda}_{ini}$ and the same scale ($s = 1$) to solutions with different $s$ values and the same $\hat{\lambda}_{fin}$. The final mass-radius curves are consistent with what is obtained by assuming an ansatz for soliton density profile \cite{Chavanis:2011mrr, Schiappacasse:2017ham} but can be obtained without making this assumption (see also \cite{Guzman:2006yc, Chavanis:2011num, Padilla:2021chr}).

An important thing to note is that the choice of $\hat{\lambda}_{fin}$ fixes only the location of a mass-radius curve in the $M-R$ plane. It is $\hat{\lambda}_{ini}$ that decides the shape of the mass-radius curve, i.e. information about what regime one is in lies with the unscaled dimensionless solutions. This is discussed in greater detail in appendix~\ref{app:regimes}. 
We note some interesting features of the mass-radius curves in presence of self-interactions: (a) For attractive self-interactions, there exists a maximum mass $M_{max}$ for $\hat{\lambda}_{ini} = -0.4$; (b) For repulsive self-interactions, for large $\hat{\lambda}_{ini} > 0$, $\hat{R}_{fin}$ appears to asymptote to a minimum radius (here the system is said to be in the Thomas-Fermi regime). These features are discussed in detail in appendix~\ref{app:extension_m_r_curves}. It is also important to note from figure~\ref{fig:m_vs_r_scaled} that for $\lambda < 0$ there exist two radii for the same soliton mass. However, only the larger of the two radii corresponds to a stable solution while the smaller one is unstable under small perturbations \cite{Chavanis:2011mrr, Chavanis:2011num, Schiappacasse:2017ham}. This establishes an upper limit on the amount of attractive self-interactions one can have if one desires a stable solitonic solution at the centre of DM halos. In terms of dimensionless self-coupling strength, this limit is given by $\hat{\lambda}_{ini} > -0.4$ (see appendix~\ref{app:extension_m_r_curves}).

\subsubsection{Implications for constraints in $\lambda-m$ plane}\label{sec:previous_work}

Having developed this machinery, we also revisit constraints from the previous work \cite{Chakrabarti:2022owq}, where the the effect of a point-mass black hole at the centre of the halo (along with SFDM self-interactions) was considered. To understand this, in appendix~\ref{app:max_mass_smbh} we present the mass-radius curve and maximum allowed mass for a fixed $\lambda < 0$ in the presence of a black hole. 

Ref.~\cite{Chakrabarti:2022owq} presented a method for obtaining observational constraints in $\lambda - m$ plane. For attractive self interactions, as is seen from figure 3 of \cite{Chakrabarti:2022owq}, there is a region of parameter space (the light grey region marked ``Can't be probed") which is such that even though the corresponding parameter values lead to a stable soliton, it can not be probed by the method presented in \cite{Chakrabarti:2022owq}. 
In appendix~\ref{app:lambda_m_BH}, we show that this inaccessible region in figure~3 of \cite{Chakrabarti:2022owq} is in-fact the region where $M_s > M_{max}$ for the corresponding values of $\lambda$ and $m$, i.e. solitonic solutions do not exist for the said $\lambda$ and $m$. Hence, there will be no region in the parameter space that corresponds to stable solitons which can not be probed by the method presented in \cite{Chakrabarti:2022owq}.

\subsection{Impact of parameters on rotation curves}\label{sec:impact_of_parameters}

Every combination of the free parameters $\{m, \hat{\lambda}_{ini}, s\}$ will lead to a unique density profile and a unique corresponding circular velocity profile. As $m$ is fixed, it is useful to ask how varying other two free parameters $\hat{\lambda}_{ini}$ and $s$, affects the velocity curves of solitons.

\begin{figure}[ht]
    \centering
    \includegraphics[width = 1\textwidth]{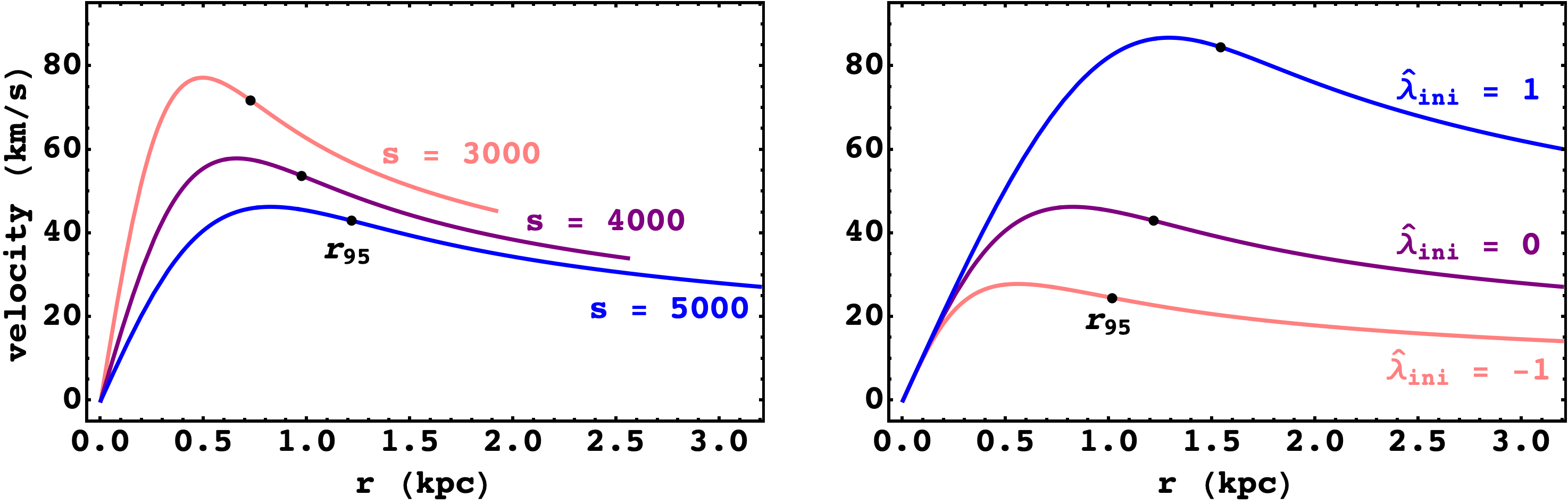}
    \caption{The left panel demonstrates how larger scale values lead to larger cores but smaller peak velocities, and vice-versa ($\hat{\lambda}_{ini} = 0$ is fixed). The right panel shows the impact of changing the self-interaction strength. For $\hat{\lambda}_{ini} > 0$, peak velocity and size of the core increases, while for $\hat{\lambda}_{ini} < 0$ peak velocity decreases and the core gets smaller ($s = 5000$ is fixed).}
    \label{fig:impact_of_parameters}
\end{figure}
\begin{enumerate} 
    \item \textbf{Impact of $s$:}
    Scaling symmetry implies that $v \rightarrow v/s$ and $r \rightarrow r\cdot s$. Therefore, an increase in $s$ leads to the stretching of the $r$-axis, while squeezing the $v$-axis, leading to a larger soliton but a smaller peak velocity. This effect is shown in the left panel of figure~\ref{fig:impact_of_parameters}. Note that $M_s$ scales in the same way as $v$, implying that for a fixed $m$ and $\lambda$ a smaller peak velocity corresponds to a lighter soliton and vice-versa.
    
    \item \textbf{Impact of $\hat{\lambda}_{ini}$:}
    As $\hat{\lambda}_{ini}$ increases, the corresponding $\hat{R}_{ini}$ and $\hat{M}_{ini}$ also increase (see blue curve in figure~\ref{fig:m_vs_r_scaled}). Hence, for a fixed $m$ and $s$, increasing $\hat{\lambda}_{ini}$ will stretch both the $r$ and $v$ axes leading to a larger peak velocity and $r_{95}$. The opposite is true when $\hat{\lambda}_{ini}$ decreases. The effect can then be described as the stretching and squeezing in roughly the direction of the slope of the linear region of the velocity curve (see right panel of figure~\ref{fig:impact_of_parameters}). Note that the squeezing effect of $\hat{\lambda}_{ini} = -1$ is much smaller than the expanding effect of $\hat{\lambda}_{ini} = 1$.
\end{enumerate}

\subsection{Confronting observed rotation curves}\label{sec:analysis_SPARC}
Armed with the knowledge of how different parameters impact soliton velocity curves, we can now confront observations. We consider low surface brightness (LSB) galaxies from the Spitzer Photometry \& Accurate Rotation Curves (SPARC) catalogue \cite{Lelli:2016zqa}, which hosts surface photometry at $3.6\ \mu m$ and HI/H$\alpha$ rotation curves for 175 galaxies. In this section, we probe the compatibility of the modified SH relation in eq.~(\ref{eq:modified_SH}) with observed rotation curves.

\subsubsection{Dataset}\label{sec:dataset}
Before proceeding, we ensure that we are dealing with good quality rotation curves by eliminating galaxies with quality flag $Q = 3$. This removes galaxies with large asymmetries and non-circular motions. Since LSB galaxies are characterized by a low effective surface brightness ($B_{\text{eff}}$), we only keep galaxies with $\log(B_{\text{eff}}) \leq 1.5\ L_\odot/\text{pc}^2$ \cite{Khelashvili:2022ffq}. This leaves us with a sample of 56 galaxies. Note that galaxies in our sample are bulgeless i.e. $V_b = 0$ in eq.~(\ref{eq:observed_vel}) at all radii for every galaxy.

\subsubsection{Rotation curves and soliton-halo relation for repulsive self-interactions}\label{sec:imposing_mod_SH_relations}

In this section, we check the compatibility of the modified SH relation in eq.~(\ref{eq:modified_SH}) with observed rotation curves of the sample of LSB dwarf galaxies from the SPARC catalogue. We carry out a procedure similar to the one performed in Ref~\cite{Bar:2021kti}, however here instead of varying scalar field mass $m$, we keep $m = 10^{-22}\ \text{eV}$ fixed and vary the dimensionless self-interaction strength $\hat{\lambda}_{ini} \geq 0$. For the sake completeness we also show the results for a varying $m$ with no self-interactions $\lambda = 0$ (for comparison with the results of \cite{Bar:2021kti}) and $\lambda \neq 0$ (in section~\ref{sec:impact_of_mass}). We also talk briefly about the effect of negative self-interactions in section~\ref{sec:why_no_att_inter}.

\begin{figure}[ht]
    \centering
    \begin{subfigure}[b]{0.45\textwidth}
        \centering
        \includegraphics[width = \textwidth]{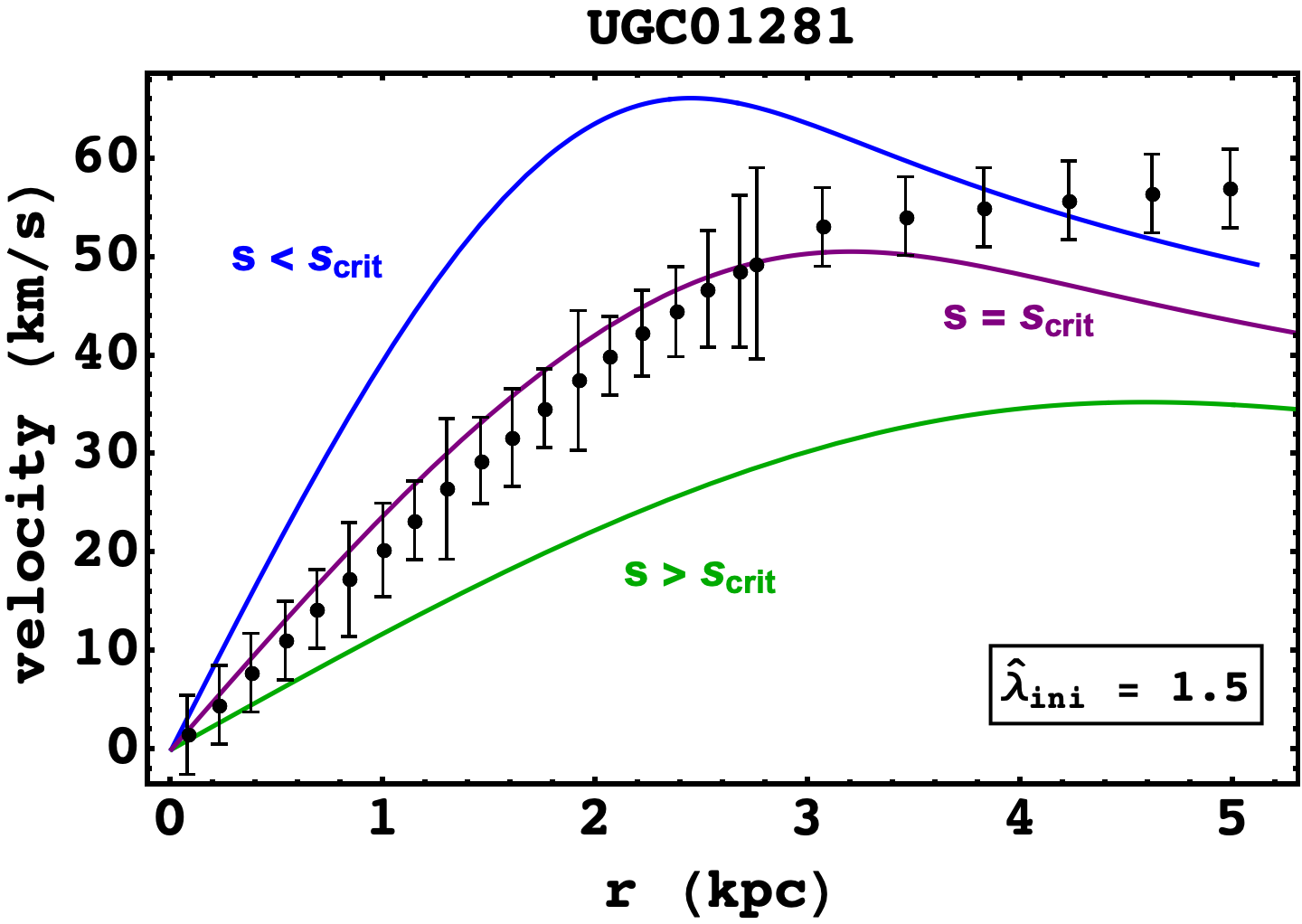}
        \caption{Black dots with error bars are the observed velocity values for `UGC 1281'. For a fixed $\hat{\lambda}_{ini} = 1.5$, as we increase $M_s$ (by decreasing $s$), the slope of the soliton-only velocity curve increases until it overshoots (blue curve) the observed velocity. Soliton masses are allowed (purple and green curves) if they fit the observed velocities or undershoot them at any radius. In case of undershooting, it is expected that the background components can be tuned to fit observations and hence are allowed.}
        \label{fig:eg_exclusion_procedure}
    \end{subfigure}
    \hfill
    \begin{subfigure}[b]{0.45\textwidth}
        \centering
        \includegraphics[width = \textwidth]{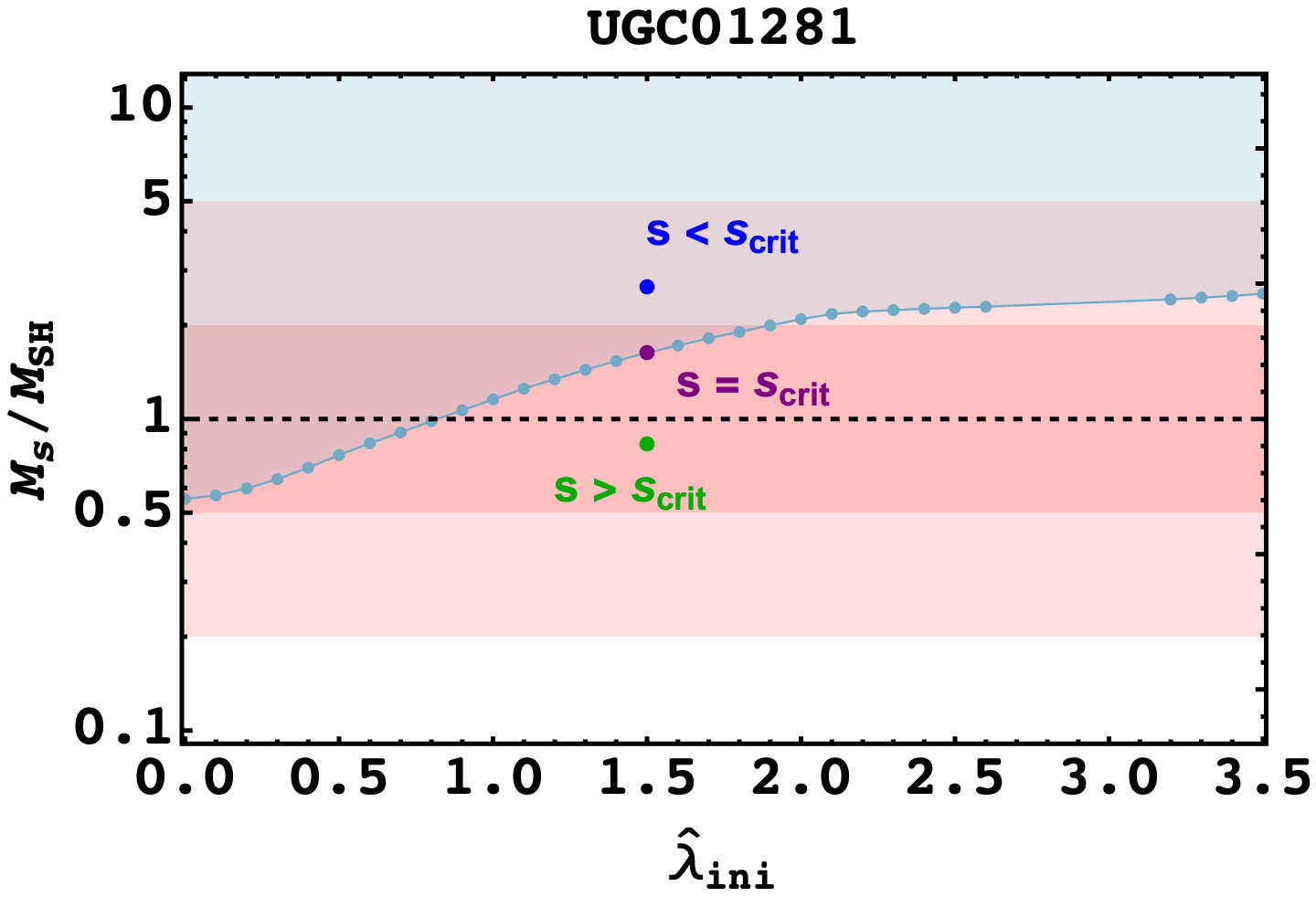}
        \caption{The maximum allowed $M_s/M_{SH}$ for various values of $\hat{\lambda}_{ini}$ are shown by the blue dots while the curve connecting them marks the boundary of the shaded (excluded) region. Solitons with $M_s/M_{SH}$ in the shaded region with lead to a velocity curve that overshoots the observed velocity. The three cases shown in figure~\ref{fig:eg_exclusion_procedure} are shown by dots of the corresponding colours. The dark and light shaded pink regions correspond to a scatter of 2 and 5 from the SH relation in eq.~(\ref{eq:modified_SH}), while the dashed line corresponds to $M_s = M_{SH}$.}
        \label{fig:eg_exclusion_region}
    \end{subfigure}
    \caption{A demonstration of the procedure detailed in this section and the resulting exclusion region for the galaxy `UGC 1281'.}
    \label{fig:eg_exclusion}
\end{figure}

To illustrate the procedure, we shall use the example of the galaxy `UGC 1281' whose observed rotation curve is shown in figure~\ref{fig:eg_exclusion_procedure} using data points. These data points correspond to $V_{obs}$ while the solid curves in figure~\ref{fig:eg_exclusion_procedure} correspond to $V_{DM}$ in eq.~(\ref{eq:observed_vel}). The dark matter velocity $V_{DM}$ can be obtained from eq.~(\ref{eq:circ_vel}) for the density profile given by eq.~(\ref{eq:tot_dens}) in which $\rho_{SFDM}$ can be obtained from the numerical solution of GPP equations. If $V_{DM}$ happens to be smaller than $V_{obs}$, the other terms on the RHS could be such that eq.~(\ref{eq:observed_vel}) is still satisfied and the corresponding model parameters leading to said $V_{DM}$ will be allowed. On the other hand, if $V_{DM}$ is larger than $V_{obs}$, eq.~(\ref{eq:observed_vel}) will not be satisfied and the corresponding model parameters will be ruled out. Note that for $r > r_t$, the density profile $\rho_{NFW}$ will be determined by NFW parameters which we assume can be adjusted to ensure that $V_{DM}$ is not larger than $V_{obs}$.

Since $m$ is fixed, free parameters of the system are $\{\hat{\lambda}_{ini}, s\}$. Now for a fixed $\hat{\lambda}_{ini}$, we have seen in figure~\ref{fig:impact_of_parameters} that a larger $s$ corresponds to a soliton velocity curve with a smaller slope in the inner region and a smaller peak velocity. Hence, for a large enough value of scale $s$ the soliton velocity curve does not overshoot the observed velocity at any point (see the green curve in figure~\ref{fig:eg_exclusion_procedure}). As $s$ decreases, the corresponding soliton mass increases (since $M_s \rightarrow M_s/s$) along with the slope of the inner region of the soliton velocity curve and its peak velocity. For a small enough value of $s$, $V_{DM}$ overshoots $V_{obs}$, as shown by the blue curve in figure~\ref{fig:eg_exclusion_procedure}. The smallest value of scale that does not cause $V_{DM}$ to overshoot $V_{obs}$ (as shown by the purple curve in figure~\ref{fig:eg_exclusion_procedure}) is denoted by $s_{crit}$. The soliton mass corresponding to $s_{crit}$ is the largest soliton mass $M_s^{crit}$ allowed by the data. For galaxies in our sample, the typical values of $s_{crit} \sim \mathcal{O}(10^4)$.

Before proceeding further, we make the following assumptions:
\begin{enumerate}
    \item We set the following overshooting condition: At an ith observed radius we calculate $\chi^2_i = \frac{(v_i^{pred} - v_i^{obs})^2}{\sigma_i^2}$, where $v_i^{pred}$ is the predicted velocity, $v_i^{obs}$ is the observed velocity and $\sigma_i$ is the uncertainty at that radius. We exclude a soliton if for any ith observed radius bin, both $v_i^{pred} > v_i^{obs}$ and $\chi_i^2 > 1$ are satisfied.\footnote{We have verified that the results shown here only change slightly when one uses $\chi^2_i > 3$ instead.}
    
    \item For $\lambda = 0$ there is an analytical expression for density profile \cite{Schive:2014hza} (also called the Schive profile) that can be evaluated at an arbitrary radius. However, for $\lambda \neq 0$ the density profile is evaluated numerically. Since the numerical solution is calculated only till the dimensionless distance $\hat{r}_{max}$, we only consider observed data points up to which a scaled soliton solution can be calculated ($r_{max} \propto m^{-1}s\hat{r}_{max}$).\footnote{However, it is expected for the density to keep falling for $\hat{r} > \hat{r}_{max}$, implying fall-off in velocity as well. Therefore if the velocity does not overshoot for any radius covered by the numerically solved part, it will not overshoot for the rest of the rotation curve as well. This is because the galaxies in our sample do not exhibit a fall-off in observed velocity at large $r$.}
    
    \item We assume that the total halo mass is the same as the CDM halo mass and does not change in the presence of a soliton at the centre \cite{Robles:2018fur, Dawoodbhoy:2021beb}. This implies that for a given galaxy, $M_h$ in eq.~(\ref{eq:modified_SH}) can be fixed to the best-fit value obtained in \cite{Li:2020iib}. 
\end{enumerate}

\vspace{0.5cm}
\noindent \textbf{\emph{Absence of self-interactions}}
\vspace{0.25cm}

\begin{figure}[ht]
    \centering
    \includegraphics[width = \textwidth]{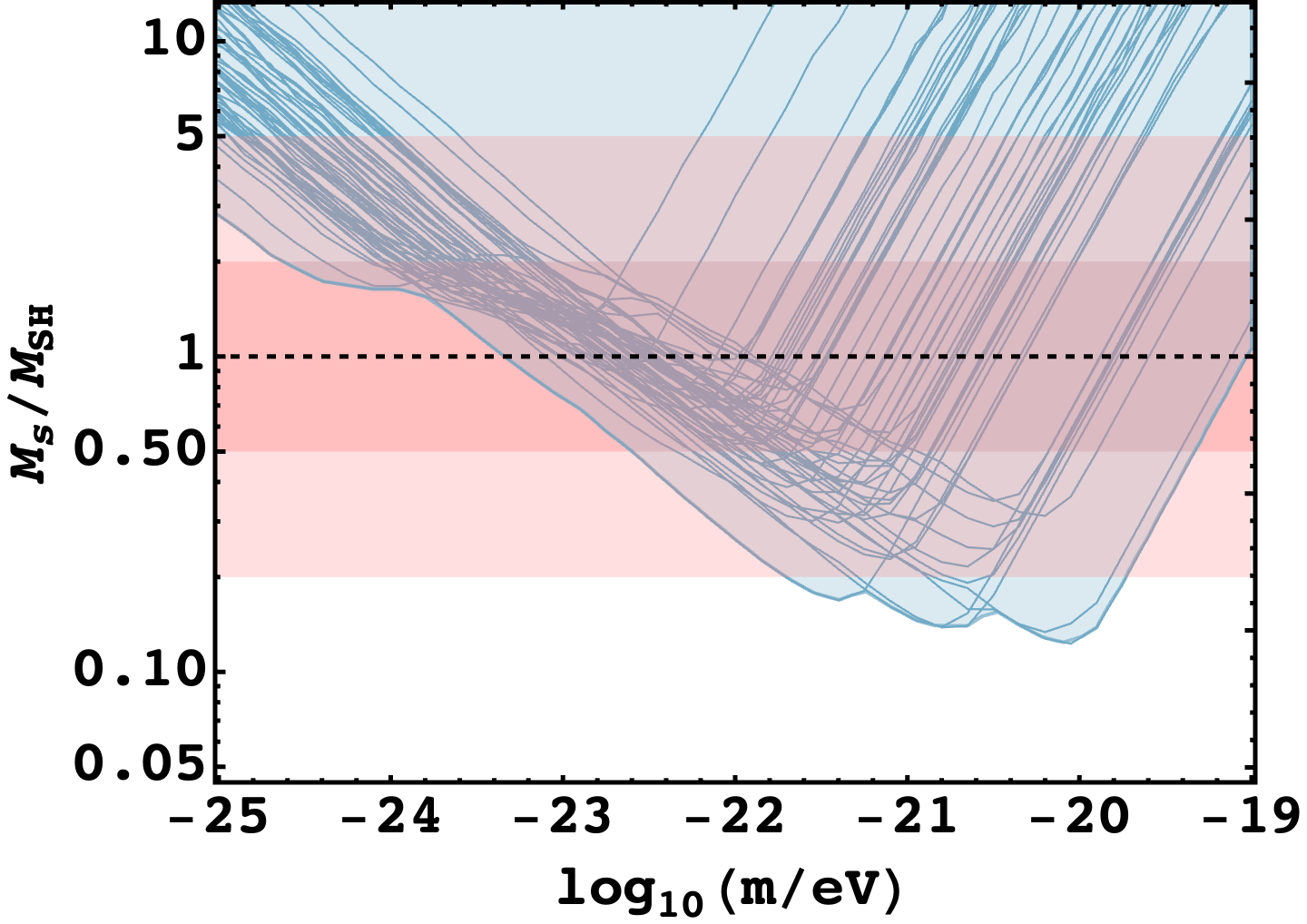}
    \caption{Here, $\hat{\lambda}_{ini} = 0$ is fixed and $m$ is allowed to vary. The dark and light shaded pink regions correspond to a scatter of factor of 2 and 5 from the SH relation in eq.~(\ref{eq:Schive_SH}) respectively.}
    \label{fig:LSB_no_inter}
\end{figure}

\noindent In the special case of SFDM with no self-interactions ($\hat{\lambda}_{ini} = 0$) i.e. FDM, the only free parameter is $s$ (when $m$ is fixed). Furthermore, soliton mass expected from soliton halo relation i.e. $M_{SH}$, is given by eq.~(\ref{eq:Schive_SH}) and hence is independent of $s$. This implies that smaller values of the ratio $M_s / M_{SH}$ corresponding to $s > s_{crit}$ will be allowed by rotation curves. Ref.~\cite{Bar:2021kti} found that for many galaxies the values of $M_s$ allowed by rotation curves are smaller than $\sim 0.5M_{SH}$ when $m$ is allowed to vary within the range $\left[10^{-24}\ \text{eV}, 10^{-20}\ \text{eV}\right]$ implying that the SH relation is not satisfied (assuming a scatter of a factor of 2 in the SH relation eq.~(\ref{eq:Schive_SH})) for solitons that are allowed by rotation curves. To verify this, we first conducted an analysis for $\hat{\lambda}_{ini} = 0$ and varied $m$ in the range $\left[10^{-25}\ \text{eV}, 10^{-19}\ \text{eV}\right]$ for the 56 LSB galaxies in our sample. The results are shown in figure~\ref{fig:LSB_no_inter}. The asymptotic dependence of $M_s/M_{SH}$ on $m$ here is consistent with what is expected from the work done in Ref.~\cite{Bar:2021kti} (i.e. $M_s/M_{SH} \propto m^{-1/2}$ for small $m$ and $M_s/M_{SH} \propto m$ for large $m$). The galaxy that imposes the strongest constraint for $m = 10^{-22}\ \text{eV}$ is `IC 2574', where all soliton masses with $M_s/M_{SH} \gtrsim 0.2$ are excluded.

However it is worth noting that while the general idea of the exclusion is the same as in Ref.~\cite{Bar:2021kti}, our approach is slightly different. For instance, as discussed earlier, overshooting is defined when $\chi^2_i$ at any single radius bin exceeds 1. Further, we also utilize the exact SH relation in eq.~(\ref{eq:modified_SH}) (which reduces to eq.~(\ref{eq:Schive_SH}) when $\hat{\lambda}_{ini} = 0$) which requires an input of $M_h$. This, along with the assumptions given earlier suggest that our results are not expected to be exactly identical.   

\vspace{0.5cm}
\noindent \textbf{\emph{Presence of repulsive self-interactions}}
\vspace{0.25cm}

\noindent Let us now investigate what happens in the presence of self-interactions. For fixed values of $M_h$, $m$ and $\hat{\lambda}_{ini}$, every value of $s$ will also correspond to a different soliton mass $M_{SH}$ expected from the SH relation in eq.~(\ref{eq:modified_SH}), since $M_{SH} \propto \sqrt{1 + As^2}$ ($A$ is just some numerical factor). As $s$ decreases, while soliton mass $M_s$ increases, $M_{SH}$ decreases, causing the ratio $M_s/M_{SH}$ to be larger. Hence for every $\hat{\lambda}_{ini}$, all $M_s/M_{SH} > M_s^{crit}/M_{SH}^{crit}$ correspond to $s < s_{crit}$. As discussed earlier, this causes $V_{DM} > V_{obs}$ and the corresponding $M_s/M_{SH}$ is excluded. The solid blue curve in figure~\ref{fig:eg_exclusion_region} is the ratio $M_s^{crit}/M_{SH}^{crit}$ for different $\hat{\lambda}_{ini}$, where the filled region above it represents the excluded soliton masses. Note that we do not expect the SH relation to be satisfied exactly. Here, the dark and light shaded pink regions in figure~\ref{fig:eg_exclusion_region} represent a scatter from eq.~(\ref{eq:modified_SH}) of factors of 2 and 5 respectively. As $\hat{\lambda}_{ini}$ increases, more of the region around $M_s = M_{SH}$ is allowed by observed rotation curves. This demonstrates that the requirement of satisfying SH relation as well as observed rotation curves can impose constraints on self-coupling of ultra-light scalar field dark matter.

The rise of the boundary curve in figure~\ref{fig:eg_exclusion_region} can be understood in the following manner: We know from the unscaled curve in figure~\ref{fig:m_vs_r_scaled} that as $\hat{\lambda}_{ini}$ increases $\hat{M}_{ini}$ also increases. Due to the shape of the inner region of the observed rotation curve, we also find that $s^{crit}$ increases with $\hat{\lambda}_{ini}$. From our earlier discussion, a larger $s^{crit}$ implies a smaller $M_s^{crit}$ and a larger $M_{SH}^{crit}$ leading to an overall smaller $M_s^{crit}/M_{SH}^{crit}$. However, the amount of increase in $\hat{M}_{ini}$ outweighs the decrease in $M_s^{crit}/M_{SH}^{crit}$ which leads to the rise of the boundary curve. 

\begin{figure}[ht]
    \centering
    \includegraphics[width = \textwidth]{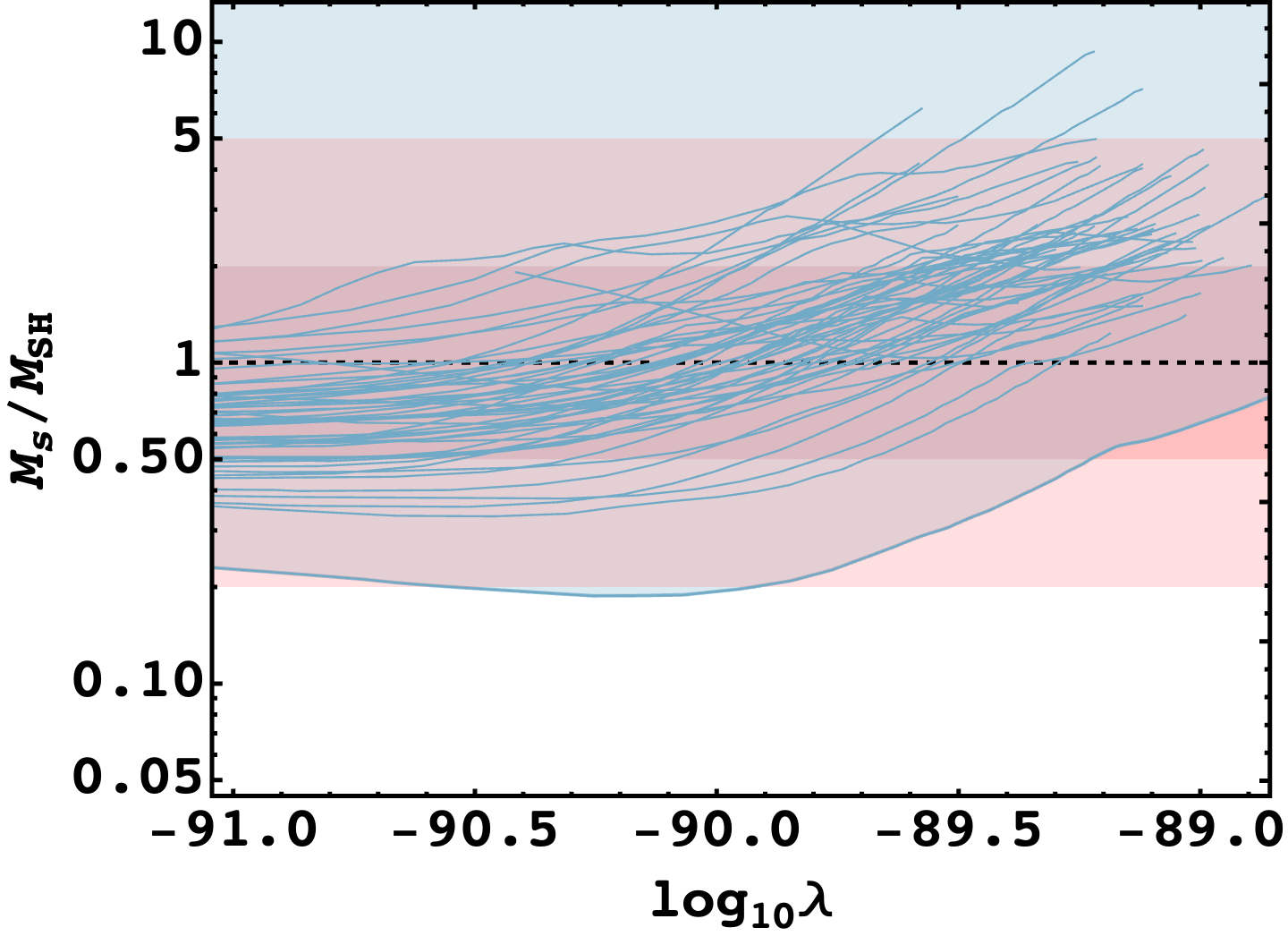}
    \caption{Here, $m = 10^{-22}\ \text{eV}$ is fixed and $\hat{\lambda}_{ini}$ is allowed to vary. The dark and light shaded pink regions correspond to a scatter of factor of 2 and 5 from eq.~(\ref{eq:modified_SH}) respectively.}
    \label{fig:LSB_with_inter}
\end{figure}

We repeat the above procedure for the remaining 55 galaxies in our sample and obtain figure~\ref{fig:LSB_with_inter} (note that in the horizontal axis we use $\lambda = 8s^2\hat{\lambda}_{ini}(M_{pl}/m)^2$). For all galaxies, the boundary between the shaded and un-shaded region is pushed upwards as $\lambda$ increases. The strongest constraints are imposed by the galaxy `IC 2574' where, for $\lambda \sim 10^{-91}$, the ratio $M^{crit}_s/M^{crit}_{SH} \sim 0.2$. As $\lambda$ increases the boundary of the excluded region is pushed upwards. This means that for large repulsive self-interactions $\lambda > \mathcal{O}(10^{-90})$, a larger region that satisfies the SH relation is allowed by rotation curves as shown in figure~\ref{fig:LSB_with_inter}. The dark and light shaded pink regions correspond to a scatter of a factor of 2 and 5 around $M_s = M_{SH}$ respectively. Note that eq.~(\ref{eq:Schive_SH}) is obtained from simulations, while eq.~(\ref{eq:modified_SH}) is its extension in the presence of self-interactions and hence we do not have an estimate for the scatter in the relation. 

Also note that some of the blue curves in figure~\ref{fig:LSB_with_inter} have a larger minimum $\lambda$. This is because the numerical solution has a finite size, requiring a larger $\hat{\lambda}_{ini}$ for the numerically calculated soliton to be large enough to reach the first observed radius bin. This leads to a larger minimum $\lambda$. On the other hand, for $\hat{\lambda}_{ini} = 3.5$ (the largest value for which we obtained a numerical solution), every galaxy will allow a different value of $s_{crit}$, leading to different values of maximum $\lambda$. For some galaxies the combination of these two effects lead to a smaller curve in figure~\ref{fig:LSB_with_inter}.

It must be noted that we have made assumptions in the beginning of this section, relaxing which could change the results of our analysis. From appendix~\ref{app:regimes}, $\hat{\lambda}_{ini} > 2.5$ implies that we are in the Thomas-Fermi (TF) regime. This is close to the values of $\hat{\lambda}_{ini}$ required to push the boundary upwards sufficiently for many galaxies. For SFDM in the TF regime (SFDM-TF), there already exist constraints on $\lambda/m^4$ from background evolution for a complex scalar field \cite{Li:2013nal, Dawoodbhoy:2021beb}. In particular, requirement of a timely transition from radiation to matter domination from CMB power spectrum imposes an upper limit on $\lambda/m^4$: For $m = 10^{-22}\ \text{eV}$ this corresponds to $\lambda \leq 10^{-89.38}$ for a real scalar field, which can further constrain the value of $\lambda$ we can allow in our analysis.

\subsubsection{Impact of attractive self-interactions}\label{sec:why_no_att_inter}

As discussed in section \ref{sec:selfinter}, axions can have attractive self interactions corresponding to a negative $\lambda$. However, as we have mentioned in section~\ref{sec:mass_radius_curves}, too strong attractive self-interactions lead to solutions that are unstable under small perturbations. The transition from stable to unstable solutions occurs at $\hat{\lambda}_{ini} = -0.4$ (see discussion in appendix~\ref{app:max_mass_no_smbh}). This sets the allowed range of $\hat{\lambda}_{ini}$ to be between 0 and -0.4. 

Also note that the second term in the square-root in the SH relation in eq.~(\ref{eq:modified_SH}) will be negative for attractive self-interactions. For a fixed value of $M_h$, a large enough $s^2\hat{\lambda}_{ini}$ will lead to an imaginary $M_{SH}$ which is unphysical. Therefore, given the small range of allowed $\hat{\lambda}_{ini}$ and the form of modified SH relation in eq.~(\ref{eq:modified_SH}), the presence of attractive self-interactions is not expected to improve constraints from the analysis carried out in this work. 

We conducted a similar analysis to the one we did for repulsive self-interactions in section~\ref{sec:imposing_mod_SH_relations} and found that within the allowed values of $\hat{\lambda}_{ini}$ the boundary of excluded region is not altered a lot for most galaxies in our sample. This is also seen from the exercise in appendix~\ref{app:slope_peak_condition} where, we compare velocity curves for different self-interaction strengths (attractive and repulsive) but the same peak velocity. From figure~\ref{fig:eg_slope_peak} note that the velocity curve for the strongest allowed attractive self-interaction strength, i.e. $\hat{\lambda}_{ini} = -0.4$ (red curve) is not very different from the velocity curve for no self-interaction case. 

\subsubsection{Impact of changing scalar field mass}\label{sec:impact_of_mass}
\begin{figure}[h]
    \centering
    \includegraphics[width = 0.5\textwidth]{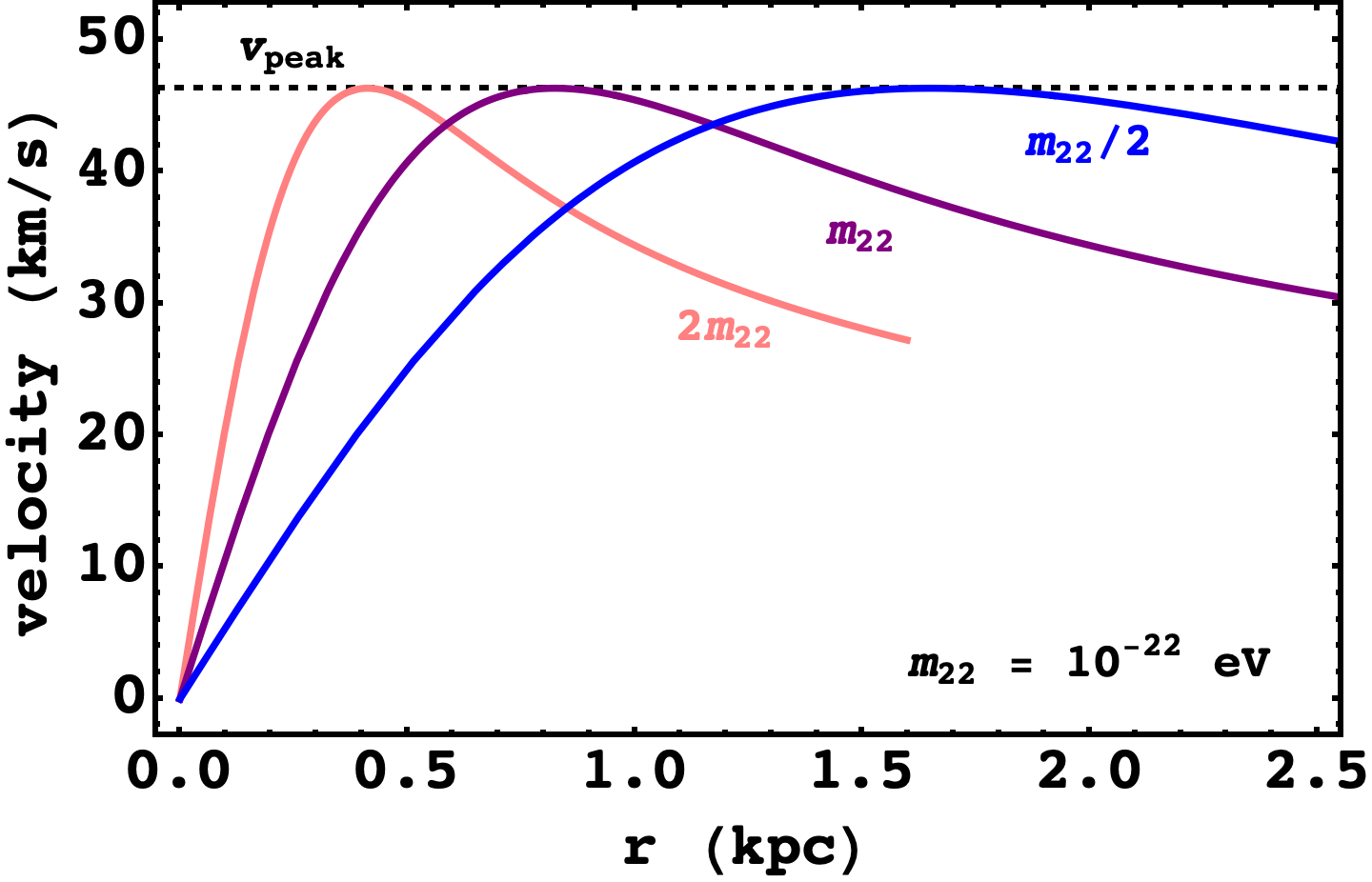}
    \caption{Velocity curves for different values of $m$ are plotted using different colours. Here values of $s = 5000$ and $\hat{\lambda}_{ini} = 0$ are fixed. The dashed horizontal lines denotes the peak velocity of each curve which remains unaffected by a change in $m$.}
    \label{fig:mass_effect}
\end{figure}

It is worth noting that in figure~\ref{fig:LSB_with_inter} we have kept $m$ fixed at its fiducial value of $10^{-22}\ \text{eV}$. It is then important to ask how changing $m$ changes (a) the velocity curves and (b) the behaviour of the boundary in figure~\ref{fig:LSB_with_inter}. The effect of changing $m$ on the velocity curves is shown in figure~\ref{fig:mass_effect}.

\begin{figure}[ht]
    \centering
    \begin{subfigure}[b]{0.45\textwidth}
        \centering
        \includegraphics[width = \textwidth]{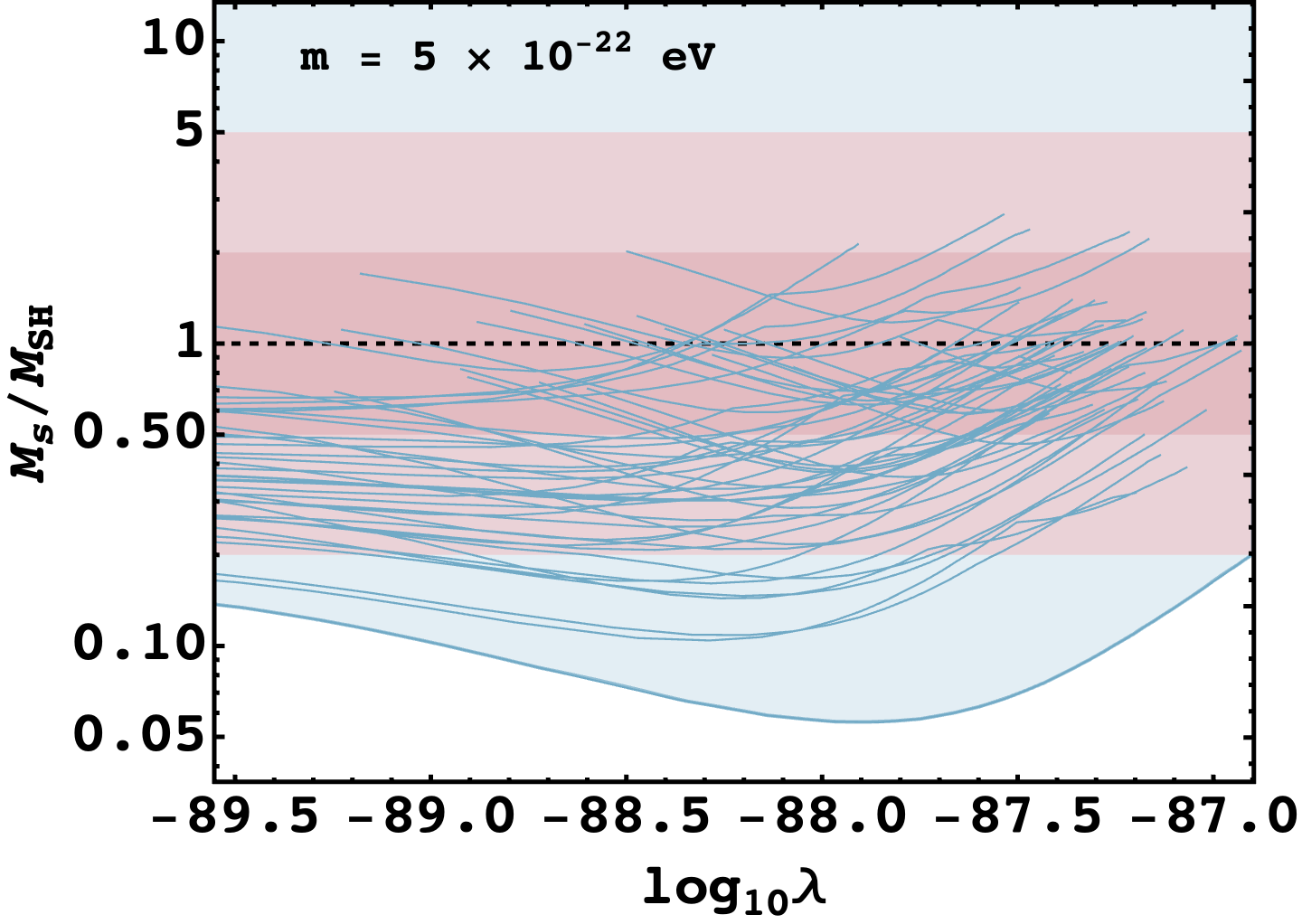}
        \caption{Here, $m = 5\times 10^{-22}\ \text{eV}$ is fixed and $\hat{\lambda}_{ini}$ is allowed to vary. The dark and light shaded pink regions correspond to a scatter of factor of 2 and 5 from eq.~(\ref{eq:Schive_SH}) respectively.}
        \label{fig:LSB_522}
    \end{subfigure}
    \hfill
    \begin{subfigure}[b]{0.45\textwidth}
        \centering
        \includegraphics[width = \textwidth]{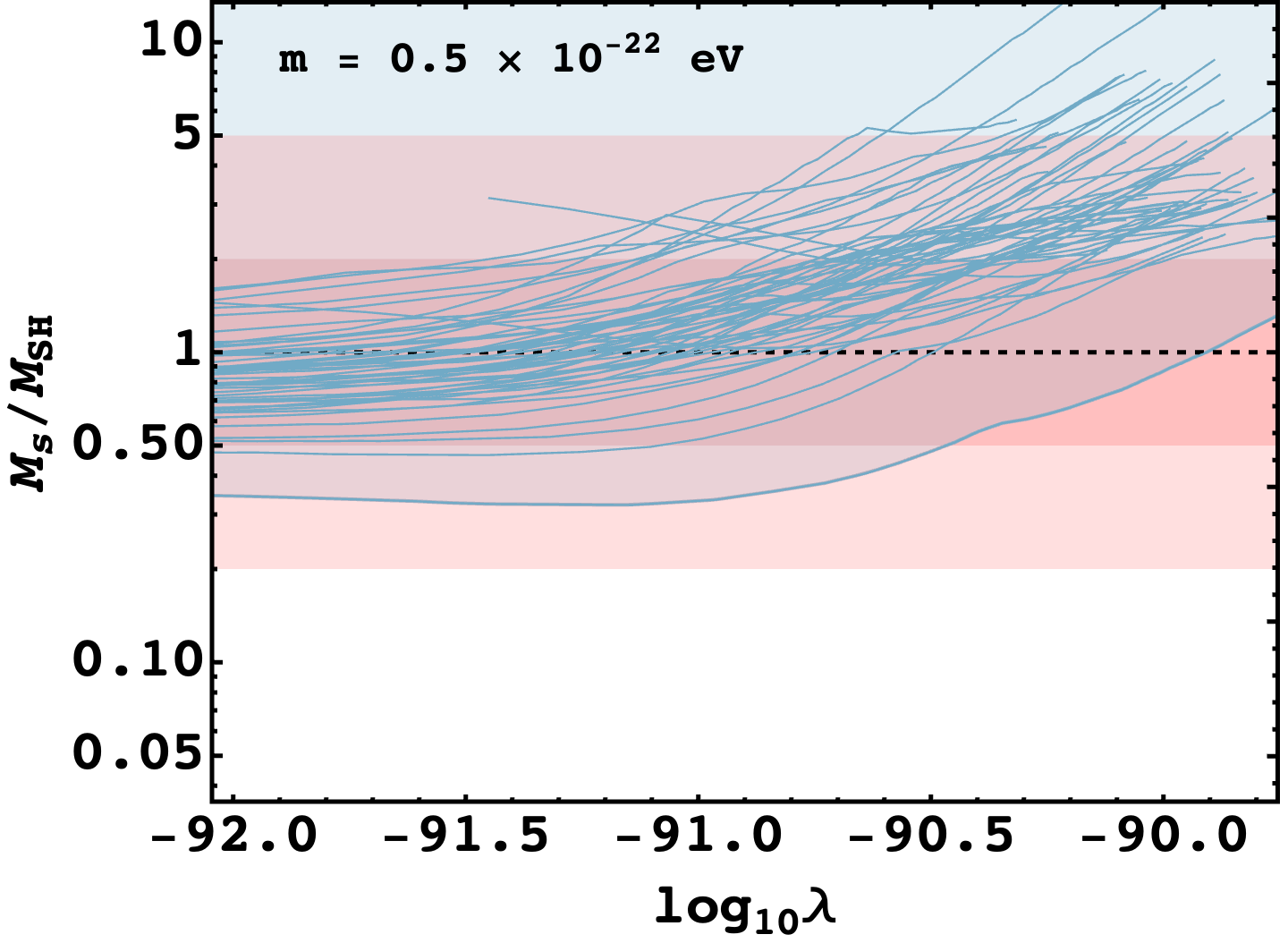}
        \caption{Here, $m = 0.5\times 10^{-22}\ \text{eV}$ is fixed and $\hat{\lambda}_{ini}$ is allowed to vary. The dark and light shaded pink regions correspond to a scatter of factor of 2 and 5 from eq.~(\ref{eq:modified_SH}) respectively.}
        \label{fig:LSB_523}
    \end{subfigure}
    \caption{Plotting $M_s^{crit}/M_{SH}^{crit}$ for a different fixed values of $m$ for 56 LSB galaxies from SPARC, where soliton masses in blue region are excluded by the data.}
    \label{fig:LSB_diff_m}
\end{figure}
From figure~\ref{fig:LSB_no_inter} it is also evident that compared to $m = 10^{-22}\ \text{eV}$, FDM masses in the range $10^{-22}$ -- $10^{-20}\ \text{eV}$ are more constrained while $m < 10^{-22}\ \text{eV}$ are less constrained by the LSB sample used. It is then important to verify whether the upwards movement of boundary occurs even when one considers a different $m$. To demonstrate this, we consider two values of scalar field mass ($m = 0.5\times 10^{-22}\ \text{eV}$ and $m = 5\times 10^{-22}\ \text{eV}$) and repeat the procedure. We find that the general behaviour of the boundary is similar to that for $m = 10^{-22}\ \text{eV}$ (see figure~\ref{fig:LSB_diff_m}). Note that since $\lambda = 64\pi\hat{\lambda}\left(\frac{m}{m_{pl}}\right)^2$, a larger value of $m$ will probe larger values of $\lambda$ for the same $\hat{\lambda}$, while the opposite is true for smaller values of $m$.

\subsubsection{Peak velocity condition}\label{sec:PVC}

Authors in \cite{Bar:2018acw} showed that for a FDM ($\lambda = 0$) core surrounded by a NFW halo, the SH relation in eq.~(\ref{eq:Schive_SH}) is equivalent to the soliton peak velocity being approximately equal to the halo peak velocity. This is also called the `velocity dispersion tracing' as seen in \cite{Chavanis:2019faf, Chavanis:2020rdo, Padilla:2021chr}. If this `peak velocity condition' (PVC) is imposed, the authors found that for $m \in \left(10^{-22}\ \text{eV}, 10^{-21}\ \text{eV}\right)$ FDM over-predicts velocities in the inner region for dark matter dominated galaxies. In other words, FDM velocity curves can either obtain the correct peak velocity or the observed slope of the inner region but cannot obtain both simultaneously. 

It is then natural to ask: Can self-interactions help? We try to answer this question in section~\ref{app:slope_peak_condition} for a sub-sample of 17 galaxies from our sample of 56. We find that for $m = 10^{-22}\ \text{eV}$ the presence of attractive self-interactions leads to more overshooting than the $\lambda = 0$ case. On the other hand for repulsive self-interactions, both PVC and the observed slope can be satisfied simultaneously. 

It is important to note that while PVC was obtained for eq.~(\ref{eq:Schive_SH}), we assume that satisfying PVC is also equivalent to satisfying the SH relation in eq.~(\ref{eq:modified_SH}). This may not be true in general since the expression for total energy of the soliton will have extra terms due to self-interactions.

However, in the absence of a SH relation, for $m = 10^{-22}\ \text{eV}$ the corresponding FDM solitons can only describe a small part of the inner regions without overshooting. On the other hand, solitons with $\lambda > 0$ can describe a large part of the inner region (sometimes the entire rotation curve) while also satisfying its observed slope.

\section{Summary}\label{sec:summary}

In order to learn about the nature of Dark Matter, the spin, mass, couplings and other fundamental properties of DM particles need to be uncovered. We considered spinless DM particles which are ultra light ($m \sim 10^{-22}$ eV). 
What could be the self-coupling strength of these particles? For axions with mass $m_a$ and decay constant $f_a$, the self coupling is suppressed and is given by $\lambda_a = - \left( \frac{m_a}{f_a} \right)^2$ and for ULAs forming dark matter, turns out to be $- {\cal O}(10^{-96})$. For other particles, the self-coupling could be positive and much larger. 

Impact of self-interacting scalar field dark matter (SFDM) on galactic rotation curves has been studied extensively in recent years. For instance in \cite{Craciun:2020twu, Harko:2022itw} authors test the SFDM model (with a complex scalar) in the TF regime against rotation curves from SPARC. They also consider additional contributions from the global rotation of the halo, random confining potentials and baryonic matter distribution. Authors in \cite{Dawoodbhoy:2021beb} also work in the TF regime. They show that an inner core described by SFDM-TF surrounded by a NFW envelope fits high mass dwarf galaxies ($M_h \sim 10^{11}\ M_\odot$) better than CDM or FDM ($\lambda = 0$) for $m = 0.8\times 10^{-22}\ \text{eV}$. On the other hand, authors in \cite{Delgado:2022vnt} use a Gaussian ansatz for $\lambda > 0$ and try to fit the inner regions of 17 bulgeless galaxies from the SPARC catalogue. They obtain a best-fit value of $\lambda \sim 2\times 10^{-90}$ and $m = 2.2\times 10^{-22}\ \text{eV}$. In this paper we did not assume a particular approximation (e.g. the TF approximation) or an ansatz to estimate density profiles. We deal directly with the numerical solutions of Gross-Pitaevskii-Poisson (GPP) equations. 

Recently \cite{Bar:2018acw, Bar:2021kti}, it has been argued that if FDM in the mass-range $10^{-24}\ \text{eV} \leq m \leq 10^{-20}\ \text{eV}$ is to be allowed by rotation curves from the SPARC database, it cannot also satisfy the soliton-halo relation in eq.~(\ref{eq:Schive_SH}) at the same time (see figure~1 of \cite{Bar:2021kti}). In section~\ref{sec:analysis_SPARC}, we obtained a similar result for a smaller sample of LSB galaxies from the SPARC database, see figure~\ref{fig:LSB_no_inter}. Later in the same section we conducted an analysis similar to the one in \cite{Bar:2021kti} but with two key differences (along with a few other minor ones): (a) Instead of varying over a range of FDM masses, we fixed $m = 10^{-22}\ \text{eV}$ and allowed the self-interaction strength $\lambda$ to vary, and, (b) we used a modified SH relation, eq.~(\ref{eq:modified_SH}), which takes into account the impact of self-interactions. We found that SFDM with $m =10^{-22}\ \text{eV}$ and $\lambda \gtrsim 10^{-90}$ can in-fact be allowed by rotation curves while simultaneously satisfying the modified SH relation within a smaller scatter than before. The upward trend of the boundary in the figure~\ref{fig:LSB_with_inter} is indicative of this effect (see section~\ref{sec:imposing_mod_SH_relations} for a detailed discussion). Note that from the analysis in appendix~\ref{app:slope_peak_condition} we found similar values of $m$ and $\lambda$ can satisfy the `peak velocity condition' within a scatter of a factor of 2 for a sub-sample of LSB galaxies. We also note that our results are in agreement with those in the appendix~E of Ref.~\cite{Chavanis:2020rdo} We briefly discussed why attractive self-interactions are not expected to play a big role in altering these constraints in section~\ref{sec:why_no_att_inter}.

We revisited the relation between the size and mass of ground state configurations of SFDM (solitons) which is sensitive to the sign and strength of $\lambda$ \cite{Chavanis:2011mrr, Schiappacasse:2017ham}. Instead of using an ansatz, we exploited the scaling symmetry to obtain the expected mass-radius curves from unscaled dimensionless numerical solutions. In appendix~\ref{app:max_mass_smbh}, we also demonstrate how the presence of a black hole at the centre alters the maximum soliton mass for a fixed $\lambda < 0$ and its impact on constraints in the $\lambda - m$ plane. In appendix~\ref{app:slope_peak_condition} we impose the criterion that the peak velocity of the soliton is approximately equal to the peak velocity of the halo and explore its implications. 

The present work motivates a full parameter search in the $\lambda-m$ parameter space which is left for future work. It will be interesting to see how constraints are altered when all 175 galaxies are taken into account, along with their baryonic contribution and parameters of the NFW envelope.

\acknowledgments
The authors would like to thank Koushik Dutta (IISER Kolkata) and Sayan Chakrabarti (IIT Guwahati) for discussions at the initial stage of the work. BD would also like to thank Manush Manju (TIFR) for help with the SPARC dataset. This work is supported by Department of Science and Technology, Government of India under Indo-Russian call for Joint Proposals (DST/INT/RUS/RSF/P-21). BD acknowledges support from the above mentioned project as a Junior Research Fellow. This research was also supported in part by the International Centre for Theoretical Sciences (ICTS) for participating in the program - Less Travelled Path to the Dark Universe (code:  ICTS/ltpdu2023/3).

\appendix

\section{Properties of solitonic solutions in the presence of self-interactions}\label{app:extension_m_r_curves}

\subsection{Different regimes of the mass-radius curves}\label{app:regimes}

In this section, we shall discuss different regimes of the mass-radius curves in the $M-R$ plane for different signs and strength of self-coupling $\lambda$. First, we shall obtain parametric dependence of $M$ and $R$ by directly comparing terms in the GPP system and then obtain numerical factors from solutions of the system. Note that we can write $\nabla = 1/R$ where $R$ is the characteristic length scale of the system. This also enables us to write the Poisson equation as $\Phi \sim 4\pi G|\psi|^2R^2$. Density can also be written as $\rho = |\psi|^2 \sim 3M/4\pi R^3$. 
\begin{enumerate}
    \item \textbf{The Non-Interacting Regime:} First we consider the case where $\lambda \rightarrow 0$. The self-interaction term is then negligible compared to other terms in eq.~(\ref{eq:GrossPitaevskii}). Comparing the gradient term and self-gravity term (and adding back missing factors of $\hbar$ and $c$) we get,
    \begin{equation}
        R \sim \frac{\hbar^2}{6GMm^2}\ .
    \end{equation}
    When $\hat{\lambda}_{ini} = 0$ there are no free parameters in the unscaled system, and there is only one solution with $\hat{M}_{ini} = 2.0612$ and $\hat{R}_{ini} = 4.822$. When scaling is introduced, $s$ becomes the sole free parameter, leading to a single mass-radius curve (solid black line in figure~(\ref{fig:m_vs_r_scaled})), where each $s$ value leads to a unique point on the curve. Using unscaled values of $\hat{M}_{ini}$ and $\hat{R}_{ini}$, eq.~(\ref{eq:scaling}), and adding back dimensions, one can write, 
    \begin{equation}\label{eq:NI_regime}
        R_{fin} = 9.94\frac{\hbar^2}{GM_{fin}m^2}\ .
    \end{equation}
    This agrees well with solutions obtained from other numerical work \cite{Chavanis:2011mrr, Chavanis:2011num}. It is straightforward to see from figure~\ref{fig:m_vs_r_scaled} that for $|\hat{\lambda}_{ini}| << 1$ the mass-radius curve for any $\hat{\lambda}_{fin}$ can be approximated by eq.~(\ref{eq:NI_regime}).

    \item \textbf{Thomas-Fermi Regime:} For $\lambda > 0$, consider the case where the outward-acting repulsive self-interactions are balanced by the inward-acting gravity such that the gradient term is negligible. From eq.~(\ref{eq:GrossPitaevskii}) 
    \begin{equation}\label{eq:TF_radius_at_home}
        R^2 \sim \frac{\lambda\hbar^3}{32\pi Gm^4c}\ = \frac{a_s\hbar^2}{Gm^3}\ ,
    \end{equation}    
    where $a_s = \frac{\lambda}{32\pi}\frac{\hbar}{mc}$ is the scattering length for a real scalar field. This is the so-called Thomas-Fermi (TF) approximation. Ref.~\cite{Boehmer:2007um} showed that in this limit, the system has an exact solution with density $\rho(r) = \rho_0 \sinc{\left(\pi r/R_{TF}\right)}$ where $R_{TF} = \pi \left(\frac{a_s\hbar^2}{Gm^3}\right)^{1/2}$. Note how $R_{TF}$ depends only on $\lambda$ (or $a_s$) and $m$. In terms of dimensionless variables, eq.~(\ref{eq:TF_radius_at_home}) can be written as $\hat{R} = \sqrt{2\hat{\lambda}_{ini}}$.\footnote{One can also write $\hat{R}_{TF} = \pi\sqrt{2\hat{\lambda}_{ini}}$.} This allows us to write the gradient term in eq.~(\ref{eq:GrossPitaevskii}) as $\hat{\nabla}^2/2 \sim \hat{R}^{-2}/2 = 1/4\hat{\lambda}_{ini}$. The TF approximation is valid when this term is negligible. We define the system to be in the TF regime when $\hat{\lambda}_{ini} \geq 2.5$ i.e. when the dimensionless gradient term in the Gross-Pitaevskii equation becomes $\mathcal{O}(0.1)$. To compare with numerical solutions, it is better to define $R_{99}$ for the analytic solution: $R_{99}^{(TF)} = 2.998 \left(\frac{a_s\hbar^2}{Gm^3}\right)^{1/2}$. An estimate for $R_{99}^{(TF)}$ can then be obtained using eqs.~(\ref{eq:lambda_scale}) and~(\ref{eq:scaling}) which in dimensionful form is  

    \begin{equation}\label{eq:R_final}
        R_{fin} = \frac{\left(\hat{R}_{ini}\hat{\lambda}_{ini}^{-1/2}\right)}{\sqrt{2}}\left(\frac{a_s\hbar^2}{Gm^3}\right)^{1/2} = 3.18\left(\frac{a_s\hbar^2}{Gm^3}\right)^{1/2}\ .
    \end{equation}
    Here we have used $\hat{\lambda}_{ini} = 3.5$ which is the largest value for which we could solve the GPP equations, with $\hat{R}_{ini} = 8.43$. This gives a good approximation of the exact value of $R_{99}^{(TF)}$.

    \item \textbf{Non-Gravitational Regime:} On the other hand, for $\lambda < 0$, when the self-interaction term dominates over the self-gravity term, the equilibrium solutions to the GPP equations are unstable under small perturbations \cite{Chavanis:2011mrr, Chavanis:2011num, Guth:2014hsa, Schiappacasse:2017ham}. Ignoring the self-gravity term in eq.~(\ref{eq:GrossPitaevskii}) and equating the gradient term with the self-interaction term, we get 

    \begin{equation}
        R \sim \frac{3\lambda\hbar M}{16\pi m^2c}\ .
    \end{equation}
    Contrary to the non-interacting regime, here $R$ and $M$ are linearly related, as shown by the left-most part of the red curve in figure~\ref{fig:m_vs_r_scaled}. In its dimensionless form, the relation simply reads $\hat{R} = 12\hat{M}\hat{\lambda}$. Note that for attractive self-interactions, the mass-radius curve transitions from the solitonic regime to the non-gravitational regime after attaining a maximum soliton mass which also separates the stable solutions from unstable ones \cite{Chavanis:2011mrr}. In the next section, we discuss the parametric dependence of this maximum mass and its numerical estimate for a fixed $\hat{\lambda}_{fin}$. 
\end{enumerate}

\subsection{Maximum mass for attractive self-interactions}\label{app:max_mass}

\subsubsection{In the absence of black hole}\label{app:max_mass_no_smbh}

There is no indication of a maximum mass when one solves the GPP equations in its unscaled dimensionless form. However, from figure~\ref{fig:m_vs_r_scaled} note that fixing the scaled $\hat{\lambda}_{fin}$ allows $\hat{M}_{fin}$ to indeed attain a maximum value. To find an expression for this maximum mass, we substitute eq.~(\ref{eq:lambda_scale}) in $M_{fin} = M_{ini}/s$,

\begin{equation}\label{eq:fin_mass}
   M_{fin} = 8\sqrt{\pi}\left(\hat{M}_{ini}|\hat{\lambda}_{ini}|^{1/2}\right)\frac{m_{pl}}{\sqrt{|\lambda_{fin}|}}\ .
\end{equation}
Since $\lambda_{fin}$ is fixed, $M_{fin} = M_{max}$ only when $\hat{M}_{ini}|\hat{\lambda}_{ini}|^{1/2}$ attains a maximum. From the unscaled numerical solutions, we find that $\left(\hat{M}_{ini}|\hat{\lambda}_{ini}|^{1/2}\right)_{max} = 0.71$, which leads to, for a real scalar field,

\begin{equation}\label{eq:max_mass}
    M_{max} \approx 10.12\frac{m_{pl}}{\sqrt{|\lambda_{fin}|}}\ .
\end{equation}
The form of eq.~(\ref{eq:max_mass}) agrees with what is obtained by \cite{Chavanis:2011mrr, Levkov:2016rkk}. It is worth noting that the maximum mass depends solely on the strength of self-interactions $\lambda$ and not on $m$. In appendix.~\ref{app:max_mass_smbh} we show how this maximum mass changes in the presence of a black hole at the centre of the halo while in appendix~\ref{app:lambda_m_BH} we discuss the implication of a different maximum mass on constraints in the $\lambda-m$ plane.

\subsubsection{In the presence of a black hole}\label{app:max_mass_smbh}
\begin{figure}[ht]
    \centering
    \begin{subfigure}[b]{0.45\textwidth}
        \centering
        \includegraphics[width = \textwidth]{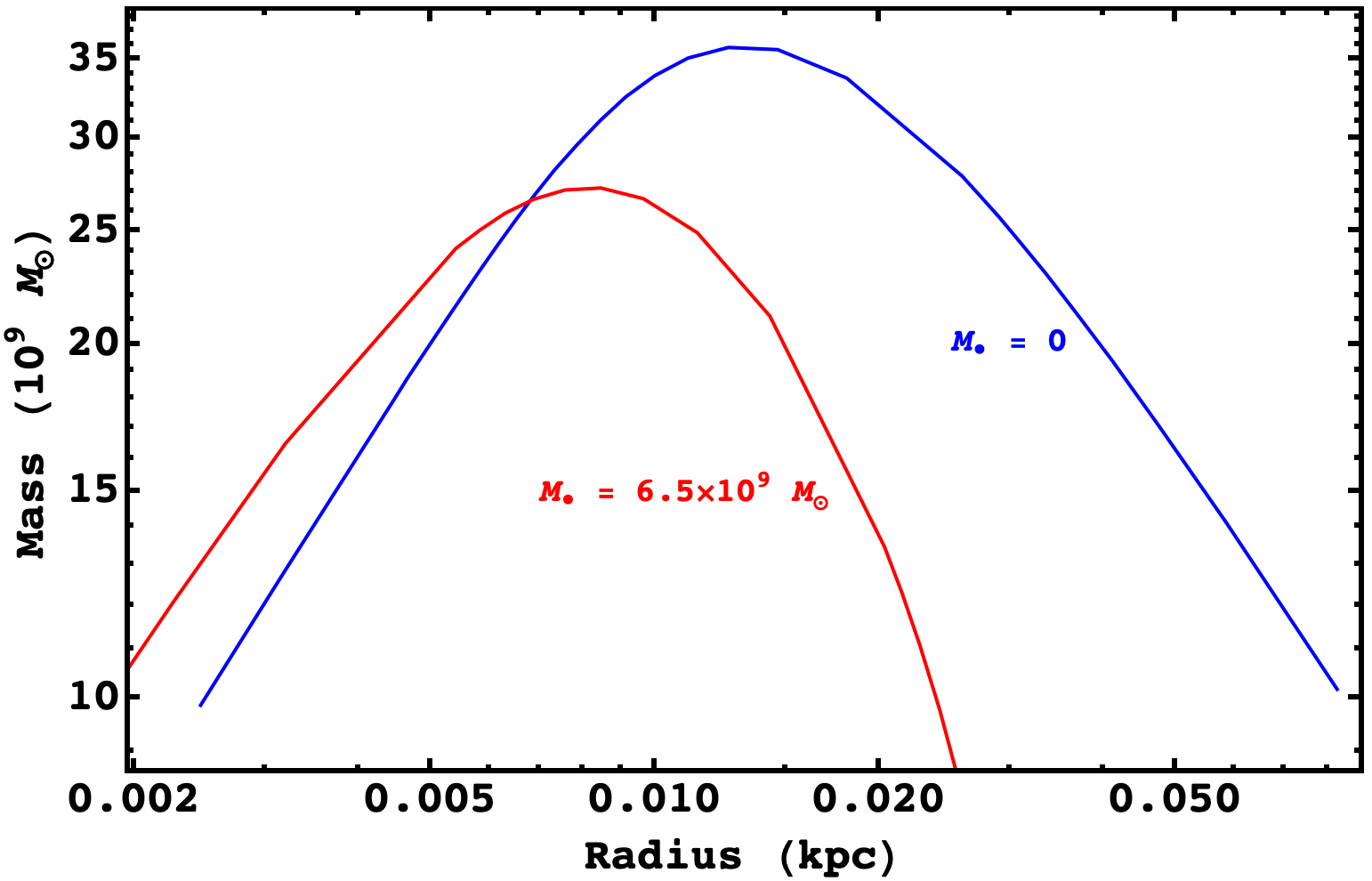}
        \caption{Effect of SMBH on maximum mass and the overall mass-radius curve for a fixed $m = 10^{-22}\ \text{eV}$ and $\lambda = 10^{-95}$.}
        \label{fig:mass_radius_smbh}
    \end{subfigure}
    \hfill
    \begin{subfigure}[b]{0.45\textwidth}
        \centering
        \includegraphics[width = \textwidth]{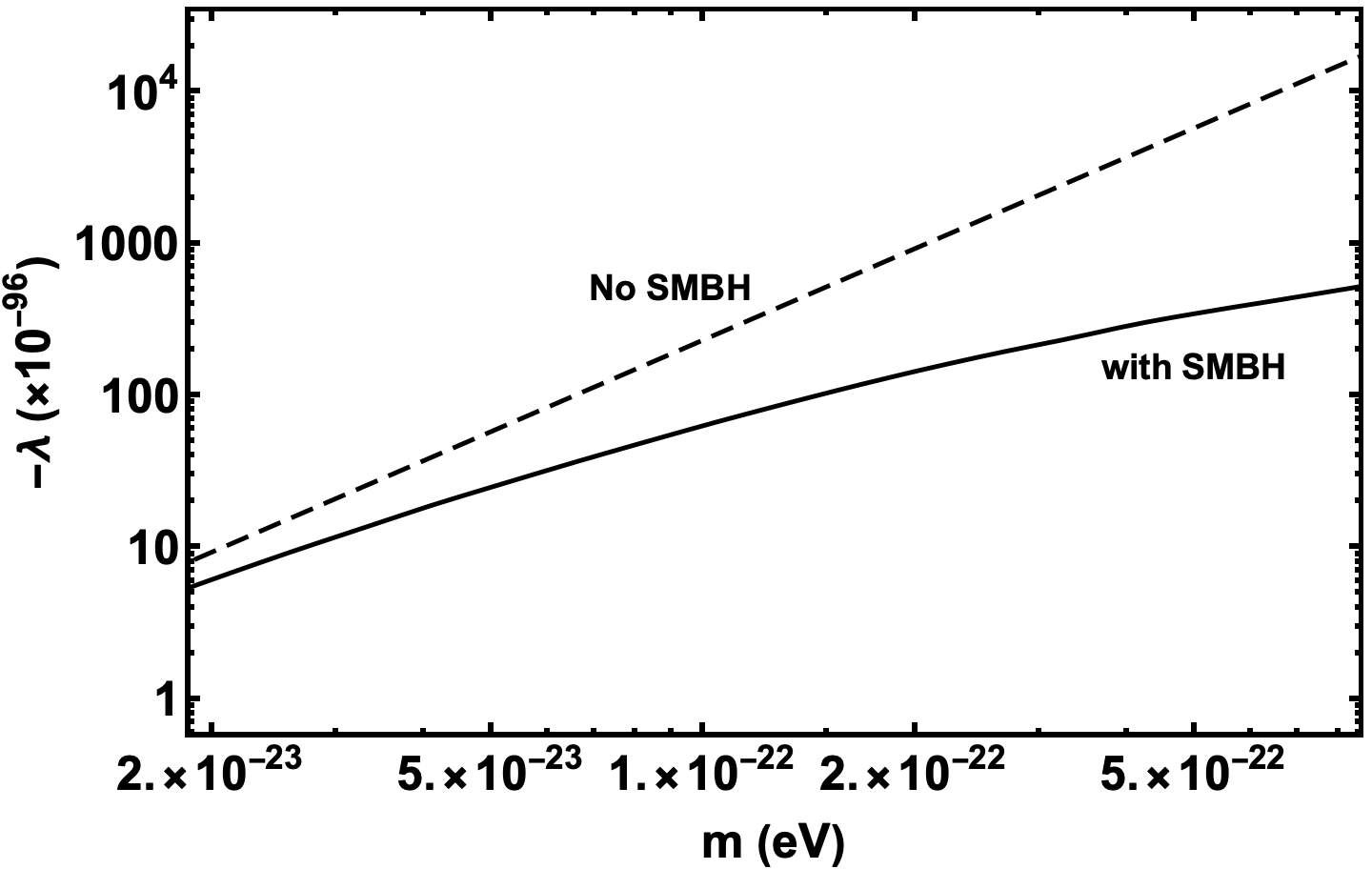}
        \caption{The maximum $\lambda$ corresponding to every $M_{fin}$ that satisfies eq.~(\ref{eq:Schive_SH}) are plotted without SMBH (dashed) and with SMBH (solid).}
        \label{fig:lambda_m_SMBH}
    \end{subfigure}
    \caption{Impact of the M87* SMBH ($M_\bullet = 6.5\times 10^9\ M_\odot$) on maximum mass in the $M-R$ plane and $\lambda-m$ is shown. Halo mass is taken to be $M_h = 2\times 10^{14}\ M_\odot$.}
    \label{fig:effect_of_SMBH}
\end{figure}

We consider a case where at the center of the halo there is a supermassive black hole (SMBH). In the Newtonian-limit we model this black hole as a point mass at the center, which leads to an extra term in the first equation of the GPP system,

\begin{eqnarray} 
 \frac{1}{2} {\hat \nabla}^2 {\hat \phi} &=& {\hat \Phi}{\hat \phi} - {\hat \gamma} {\hat \phi} - \frac{\hat \alpha}{\hat r} {\hat \phi} + 2 {\hat \lambda} {\hat \phi}^3  \; , \label{eq:GP_dimless_smbh}\\
  {\hat \nabla}^2 {\hat \Phi} &=&{\hat \phi}^2  \; .\label{eq:P_dimless_smbh}
\end{eqnarray}
Here $\hat{\alpha} \equiv GM_\bullet m/\hbar c$ and $M_\bullet$ is the mass of the black hole. Addition of a Newtonian point mass term keeps the scaling symmetry intact \cite{Davies:2019wgi, Chakrabarti:2022owq} and for $\hat{\alpha}$ it is found to be ($\hat{\alpha} \rightarrow \hat{\alpha}/s$). Similar to the discussion in section~\ref{sec:mass_radius_curves}, to determine maximum mass in the presence of attractive self-interactions, we first fix $\hat{\lambda}_{fin}$ to determine scale $s$. Since the physical mass of the SMBH $M_\bullet$ is also fixed, we also fix some $\hat{\alpha}_{fin}$ such that $\hat{\alpha}_{ini}/s = \hat{\alpha}_{fin}$ for every $\hat{\lambda}_{ini}$. We incorporate this in the following manner: 

\begin{enumerate}
    \item For a fixed $m$ and $M_\bullet$ find the desired $\hat{\alpha}_{fin}$.
    \item Fix $\hat{\lambda}_{fin}$ such that for every $\hat{\lambda}_{ini}$ one gets the required scale $s$ from eq.~(\ref{eq:lambda_scale}).
    \item Using scaling symmetry, find $\hat{\alpha}_{ini} = s\hat{\alpha}_{fin}$, for every $\hat{\lambda}_{ini}$.
    \item Solve the GPP system to obtain $\hat{M}_{ini}$ and $\hat{R}_{ini}$ for the given $\hat{\lambda}_{ini}$ and $\hat{\alpha}_{ini}$. 
\end{enumerate}
Because of the squeezing effect of $\hat{\alpha}_{ini}$ \cite{Davies:2019wgi} on the density profile, the corresponding $\hat{M}_{ini}$ and $\hat{R}_{ini}$ will be different compared to $\hat{\alpha}_{ini} = 0$. It is then expected that $M_{max}$ will be altered in the presence of a black hole. Using the same argument as in appendix~\ref{app:max_mass_no_smbh} and eq.(\ref{eq:max_mass}) maximum mass for a fixed $\lambda_{fin}$ in the presence of a black hole can be determined by $(\hat{M}_{ini}|\hat{\lambda}_{ini}|^{1/2})_{max}$. As an example, we consider the M87 halo with $M_h = 2\times 10^{14}\ M_\odot$ and $M_\bullet = 6.5\times 10^9\ M_\odot$. For $m = 10^{-22}\ \text{ev}$ the desired scaled $\hat{\alpha_{fin}} = 0.0048$. We plot the mass-radius curve for both $\hat{\alpha} = 0$ (blue curve) and $\hat{\alpha}_{fin} = 0.0048$ (see red curve) in figure.~\ref{fig:mass_radius_smbh}. We find that $M_{max}$ is smaller in the presence of a black hole, which is in agreement with the results in \cite{Chavanis:2019bnu}.

\subsection{$\lambda$-$m$ plane for $\hat{\lambda}_{ini} < 0$}\label{app:lambda_m_BH}

From the previous discussion note that $\hat{\alpha}_{ini}$ is dependent on $m$ through $\hat{\alpha}_{fin}$. One can therefore get a corresponding curve in the $\lambda-m$ plane by fixing the scaled soliton mass $\hat{M}_{fin}$ instead of $\hat{\lambda}_{fin}$, setting the desired scale to be $s = \hat{M}_{ini}/\hat{M}_{fin}$ for every value of $\hat{\lambda}_{ini}$. This was the procedure adopted by authors in \cite{Chakrabarti:2022owq} and it transforms eq.~(\ref{eq:max_mass}) into

\begin{equation}
    |\lambda_{max}| = 64\pi \left(\hat{M}_{ini}|\hat{\lambda}_{ini}|^{1/2}\right)_{max}^2\frac{m_{pl}^2}{M_{fin}^2}\ .
\end{equation}
Now for every soliton mass $M_{fin}$ there is a maximum allowed $\lambda_{max}$ such that $M_{fin} = M_{max}(\lambda_{max})$. In the presence of black hole, it is then clear that if $M_{fin}$ remains unchanged, $\lambda_{max}$ will be smaller. One can fix the scaled soliton mass by assuming that $M_{fin}$ must satisfy some soliton-halo relation (e.g. eq.~(\ref{eq:Schive_SH})) for a fixed $M_h$. The resultant curve in the $\log{\lambda}-\log{m}$ plane when $\hat{\alpha}_{fin} = 0$ is a straight line (dashed curve in figure~\ref{app:lambda_m_BH}). However, $\hat{\alpha}_{fin} \neq 0$, for every $\hat{\lambda}_{ini}$ and $m$ one must find some $\hat{\alpha}_{ini}$ such that $\hat{\alpha}_{fin}$ as well as $\hat{M}_{fin}$ correspond to the desired $M_\bullet$ and soliton mass $M_{fin}$ respectively. Hence, for every $m$ the corresponding $\lambda_{max}$ will be smaller than the case with $\hat{\alpha}_{fin} = 0$. as shown by the solid curve in figure~\ref{fig:lambda_m_SMBH}.

In figure~3 of \cite{Chakrabarti:2022owq}, it was noted that the light gray region labeled `cannot be probed' was the region in $\lambda - m$ plane where there could exist solitonic solutions but were not accessible by the method proposed in the paper. However, from the previous discussion, the `cannot be probed' is the same as the `not allowed' region (where $\lambda > \lambda_{max}$ and no solitonic solutions exist) that has been distorted and pushed down due to the presence of a black hole at the centre. A detailed discussion regarding the behaviour of the curve can be found in the appendix~C of \cite{Chakrabarti:2022owq}.

\section{Satisfying both observed slope and peak velocity}\label{app:slope_peak_condition}

In this section we shall impose the `peak velocity condition' (PVC) as a proxy for SH relation in eq.~(\ref{eq:Schive_SH}) \cite{Bar:2018acw} which states that peak velocity in the halo should be roughly equal to the peak velocity in the soliton. Without obtaining halo fits, one can estimate halo peak velocity for a galaxy as the DM velocity ($V_{DM}$) measured in the flat region of the rotation curve sufficiently far from the centre. We estimate this as $V_{DM}$ obtained at the farthest observed radius bin \cite{Bar:2018acw}.

For this section, in addition to the LSB condition, we also consider dwarf galaxies with halo masses $M_h \sim 10^{10}\ M_\odot$, with maximum velocities $v_{max} \sim 80\ \text{km/s}$ bringing our number down to 36 galaxies. This constraint comes from the limitation of our numerical approach where solutions for $\hat{\lambda}_{ini} > 3.5$ are difficult to obtain due to the fine-tuned initial guesses required for $\hat{\gamma}$. We further note that not all galaxies in the sample have flat rotation curves at farthest measured radii, i.e. there are no measurements far enough from the center to establish a halo peak velocity. Hence, we shall discard galaxies with no established $V_{flat}$ (provided by the SPARC catalogue) for this analysis, leaving us with 17 galaxies. We first fix $m = 10^{-22}\ \text{eV}$ to focus on the effect of self-interactions. We vary $\{\hat{\lambda}_{ini}, s\}$ and minimize reduced $\chi^2$ \footnote{Reduced $\chi^2$ is defined as $\chi^2_{\text{red}} = \frac{1}{N-K}\sum_i^N\frac{(v_i^{pred} - v_i^{pred})^2}{\sigma_i^2}$ where $N$ is the number of data points and $K$ is the number of free parameters.} for all 17 galaxies while also requiring that the peak velocity condition is satisfied within some scatter. First, we make the following assumptions:

\begin{enumerate}    
    \item SPARC database also provides contribution of baryonic components like gas, disk and bulge to the total observed velocity for every galaxy. Using the best fit value $\Upsilon_d = 0.5$ \cite{Lelli:2016zqa}, we obtain DM-only rotation curves $V_{DM}(r)$ from eq.~(\ref{eq:observed_vel}). 

    \item We are only interested in whether SFDM can describe inner regions of galaxies and hence do not try to fit the NFW envelope. We assume that at some $r = r_t$, the NFW profile takes over and free parameters $\{r_t, r_s\}$ can be adjusted to fit the rest of the rotation curve. 

    \item We also assume that $r_t = r_{95}$, ensuring that NFW envelope takes over sufficiently far from the radius ($r_p$) at which the soliton velocity peaks (see figure~\ref{fig:density_velocity}). Hence, we consider data-points where the corresponding radius bin $r_{obs} \leq r_{95}$. Given that we have two free parameters $\{\hat{\lambda}_{ini}, s\}$, we only allow parameters for which the number of data-points within $r_{95}$ is at least 3. 

    \item Soliton and halo peak velocities are denoted by $v_p^{(s)}$ and $v_p^{(h)}$ respectively. We allow for a scatter of a factor of 2 from $v_p^{(s)} = v_p^{(h)}$ which is the allowed scatter in eq.~(\ref{eq:Schive_SH}) \cite{Schive:2014hza, Bar:2018acw}. Hence, only parameters satisfying $0.5 < v_p^{(s)}/v_p^{(h)} < 2$ are considered.
\end{enumerate}
To illustrate the effect of imposing PVC on SFDM solitons with attractive or repulsive self-interactions, we first consider the velocity curve for the galaxy `UGC 1281' for which $v_p^{(h)} = 51.5\ \text{km/s}$.

\begin{figure}[ht]
    \centering
    \includegraphics[width = 0.6\textwidth]{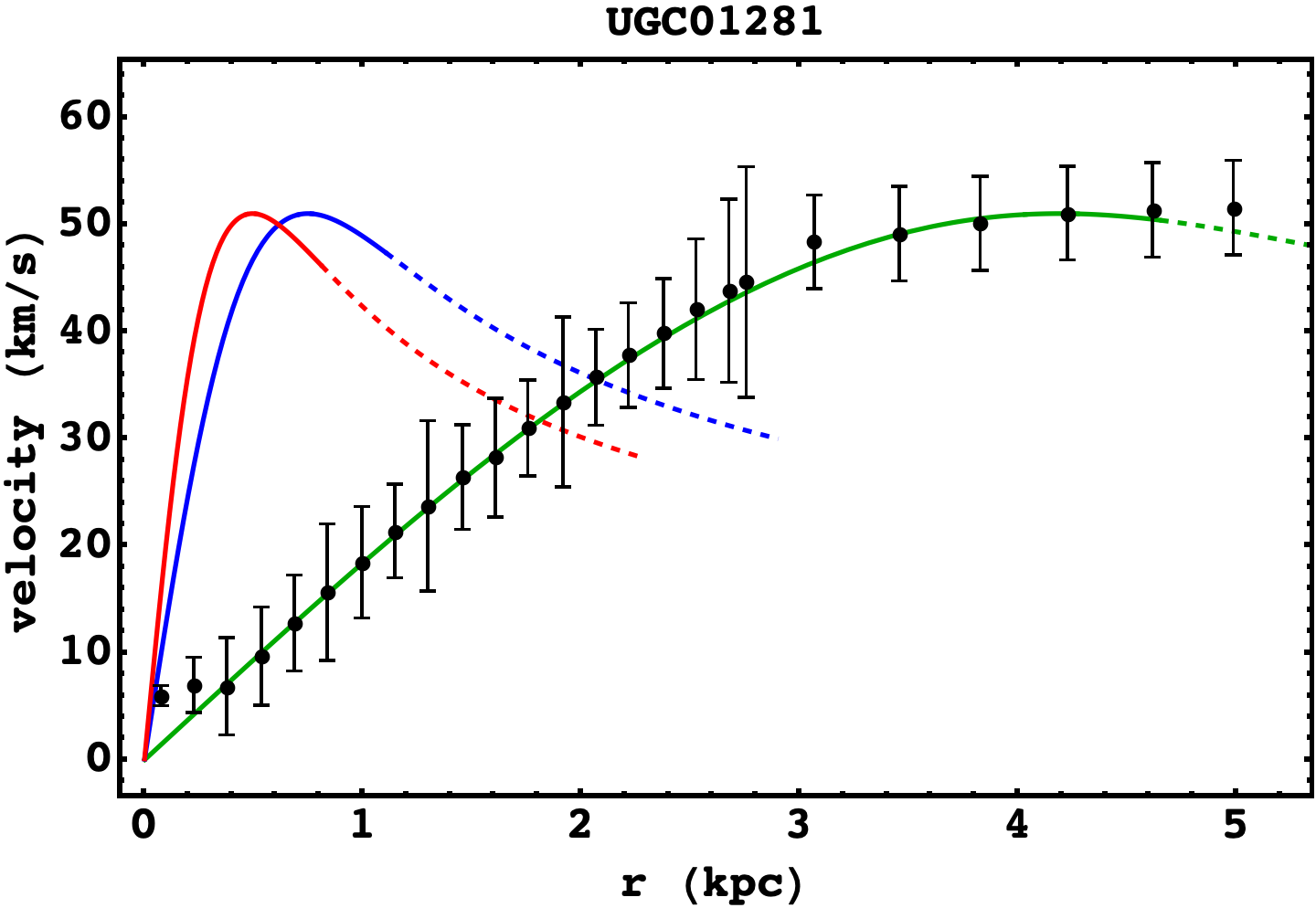}
    \caption{The green curve represents a soliton with $\hat{\lambda}_{ini} = 2$ while the blue and red curves correspond to $\hat{\lambda}_{ini} = 0$ and $\hat{\lambda}_{ini} = -0.4$ respectively. Dashed curves correspond to $r > r_{95}$. The corresponding scaled self-coupling strength is $\lambda = 3.98\times 10^{-90}$ for repulsive self-interactions.}
    \label{fig:eg_slope_peak}
\end{figure}

\begin{figure}[ht]
    \centering
    \includegraphics[width = \textwidth]{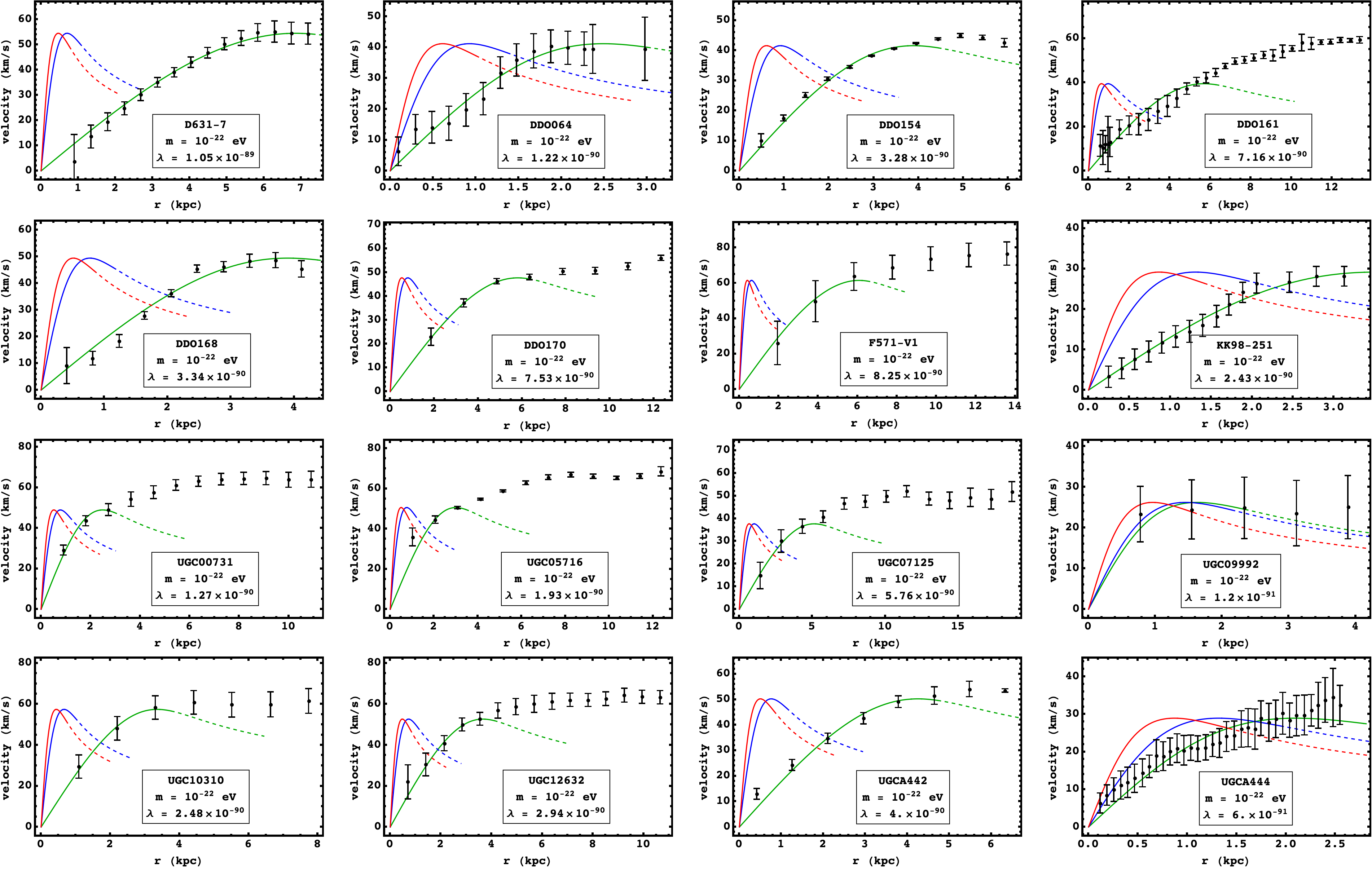}
    \caption{Remaining 16 galaxies from LSB galaxy sample. Blue curves represent solitons with no self-interactions, while red and green are solitons with attractive ($\hat{\lambda}_{ini} = -0.4$) and repulsive self-interactions respectively. For all curves, $50\%$ deviation from PVC is allowed. The dashed part of the soliton curves represents $r > r_{95}$. See text in appendix~\ref{app:slope_peak_condition} for discussion.}
    \label{fig:LSB_PVC}
\end{figure}

\begin{enumerate}
    \item \textbf{No Self-Interactions:} The blue curve in figure~\ref{fig:eg_slope_peak} corresponds to the velocity curve of a soliton with $\lambda = 0$ and $m = 10^{-22}\ \text{eV}$. Here the only free parameter is the scale $s$, which is chosen to satisfy $v_p^{(s)} = 51\ \text{km/s} = v_p^{(h)}$. The slope of the resultant soliton velocity curve is quite steeper than the slope of the observed curve. From the discussion in section~\ref{sec:impact_of_parameters}, it is clear that if one tries to reduce the slope by increasing $s$, peak velocity will consequently be smaller. Hence, for $\lambda = 0$, one cannot satisfy the observed slope and peak simultaneously, which is in agreement with the results of \cite{Bar:2018acw}.

    \item \textbf{Attractive Self-Interactions:} Now, we consider $\lambda < 0$ and allow $\hat{\lambda}_{ini}$ to vary. For every $\hat{\lambda}_{ini}$ one can choose a scale $s$ such that $v_p^{(s)} \approx v_p^{(h)}$. However, since we require a stable soliton, we cannot consider any $\hat{\lambda}_{ini} < -0.4$ (see section~\ref{sec:mass_radius_curves}), limiting the effect of attractive self-interactions. Moreover, as discussed in section~\ref{sec:impact_of_parameters} presence of attractive self-interactions squeezes the velocity curve in the direction of the slope requiring a smaller $s$ value to attain the same soliton peak velocity as the $\lambda = 0$ case. This leads to an even steeper slope, worsening the prediction (see red curve in figure~\ref{fig:eg_slope_peak}).
    
    \item \textbf{Repulsive Self-Interactions:} Finally, we allow $\hat{\lambda}_{ini} > 0$ to vary and find that a large $\hat{\lambda}_{ini} = 2$ allows one to satisfy the peak velocity condition without over-predicting velocities in the inner region. One can choose an $s$ to satisfy the observed slope in the inner region, while $\hat{\lambda}_{ini}$ can be varied to obtain the correct peak velocity. This is because a larger $\hat{\lambda}_{ini}$ will lead to a larger soliton, which alters the peak velocity while keeping the slope of the inner region roughly unchanged (see section~\ref{sec:impact_of_parameters}). For `UGC 1281' $\hat{\lambda}_{ini} = 2$ and $s = 11942$ satisfies the observed slope and peak simultaneously. The corresponding scaled value of self-coupling is $\lambda = s^2\lambda_{ini} = 3.98\times 10^{-90}$. 
\end{enumerate}
Velocity curves corresponding to parameters with minimum $\chi^2_{\text{red}}$ for the remaining 16 galaxies are shown in figure~\ref{fig:LSB_PVC} by the green curves. Note that dimensionless self-coupling $\hat{\lambda}_{ini}$ is allowed to vary within the range $\left[-0.4, 3.5\right]$ ensuring stable solutions (for attractive self-interactions).

For all galaxies, SFDM with repulsive self-interactions better describes the slopes of inner regions. Curves for attractive self-interactions using $\hat{\lambda}_{ini} = -0.4$ (red) and no self-interactions (blue) with the same soliton peak velocity as the best-fit case are also shown for comparison. `DDO 161' exhibits the highest deviation from PVC ($v_p^{(s)} \sim v_p^{(h)}/2$) for the least $\chi^2_\nu$ value, while describing only a part of the linearly increasing region. On the other hand, in many cases velocity curves of solitons with $\hat{\lambda}_{ini} > 0$ trace out a large portion of the observed rotation curve. Attractive self-interactions however, already constrained to be small to ensure stability, do not appear to be well probed by the rotation curve data presented here. Even with their limited effect on velocity curves, they seem to fare worse, predicting steeper slopes than even the no self-interactions case for the same peak velocity. 

For repulsive self-interactions, we find that large values of self-coupling ($\hat{\lambda}_{ini} \geq 1$) are preferred for many galaxies. The values of $\hat{\lambda}_{ini}$ seems to suggest that the solitons are close to the Thomas-Fermi regime (see appendix~\ref{app:regimes}). However, given the diversity of slopes and sizes of inner regions, it appears to be unlikely that a single value of $\lambda$ will fit all rotation curves. The order of magnitude of self-interaction strength is $\mathcal{O}(10^{-90})$, with an average value $\lambda \sim 3.92\times 10^{-90}$. We stress that we do not attempt to conduct a full parameter estimation and perform only a crude analysis to study the effects of self-interactions for a fixed $m = 10^{-22}\ \text{eV}$. An important caveat here is that we do not allow for the baryonic contribution to change ($\Upsilon_d = 0.5$ is fixed) or fit the outer envelope using the NFW profile. One can potentially tune these components appropriately to obtain best fits for a fixed $\lambda$, which is left for future work.
\end{document}